\documentclass[aps,twocolumn]{revtex4-1}
\usepackage{amsfonts}
\usepackage{graphicx}
\usepackage{dcolumn}
\usepackage{bm}
\usepackage{amsmath}
\usepackage{epsfig}
\usepackage{amssymb}
\usepackage{times}

\begin{document}
\title{Sub-Gaussian and subexponential fluctuation-response inequalities}

\author{Yan Wang}
\email{wangyan@iastate.edu; ywang197@gmail.com}
\affiliation{Department of Statistics, Iowa State University, Ames, Iowa 50011, USA}

\begin{abstract}
  Sub-Gaussian and subexponential distributions are introduced and applied to study the fluctuation-response relation out of equilibrium. A bound on the difference in expected values of an arbitrary sub-Gaussian or subexponential physical quantity is established in terms of its sub-Gaussian or subexponential norm. Based on that, we find that the entropy difference between two states is bounded by the energy fluctuation in these states. Moreover, we obtain generalized versions of the thermodynamic uncertainty relation in different regimes. Operational issues concerning the application of our results in an experimental setting are also addressed, and nonasymptotic bounds on the errors incurred by using the sample mean instead of the expected value in our fluctuation-response inequalities are derived.
\end{abstract}

\maketitle

\section{Introduction}\label{sec:intro}

Gaussian distributions play a crucial role in statistical physics. As a classical example, when a thermodynamical system is at equilibrium, a typical physical quantity fluctuates around its ensemble average in a Gaussian way \cite{Book:Landau}. Two amazing properties of a Gaussian random variable are that (i) it can be fully characterized by its mean and variance, without resorting to higher order statistics; (ii) the Gaussian property is preserved under linear transformation. Thanks to such properties, the highly successful linear response theory was established \cite{Book:Kubo,PhysRep:2008}, which relies heavily on the assumption that when the perturbation is weak, the deviation from the original equilibrium state is small, and one can work in the regime where the leading effect of an external force is linear. However, when the perturbation is strong and the nonlinear effect has to be taken into account, the linear response theory is no longer working well. Recent years have witnessed substantial advances in nonequilibrium statistical physics. Results such as the Jarzynski equality \cite{PRL:Jarzynski} and various kinds of fluctuation relations \cite{RPP:Seifert} have been shown to be valid for quite general nonequilibrium processes, beyond the linear response regime. Such theories consider the system's evolution path in phase space, and associate each trajectory with some physical quantities like (stochastic) work and entropy production that are defined in an unusual sense. In an experiment that manipulates a single molecule of RNA between two conformations, it is found that, when the perturbation is weak, the distribution of trajectory-dependent dissipated work can be well approximated by a Gaussian distribution, but this is not true for stronger perturbations \cite{Science:2002}. Actually, a large body of literature exists regarding the non-Gaussian distributions encountered in nonequilibrium physics, both theoretically and experimentally; for example, see Refs. \cite{Nature:2005,PRL:VdB2006,PRL:Seifert2006,PRL:2009,PRE:2009,arXiv:H2020}.

Despite being non-Gaussian, it seems that these distributions typically are unimodal, and they differ from a Gaussian mainly in the existence of skewness and a different decay rate in the tail probability. One might correct such deviations by considering higher order statistics \cite{PRL:VdB2006,PRL:2019,arXiv:Q2020}, but that may require the dynamics of the system, or the ``perturbation'' may be too strong to be treated in a perturbative way. In this work, we introduce the sub-Gaussian and subexponential distributions as two classes of distribution that are particularly relevant to nonequilibrium physics. In particular, concrete examples are provided regarding their relevance to the Jarzynski equality and the thermodynamic uncertainty relation.

\begin{figure}[tbp]
  \begin{center}
  \epsfig{figure=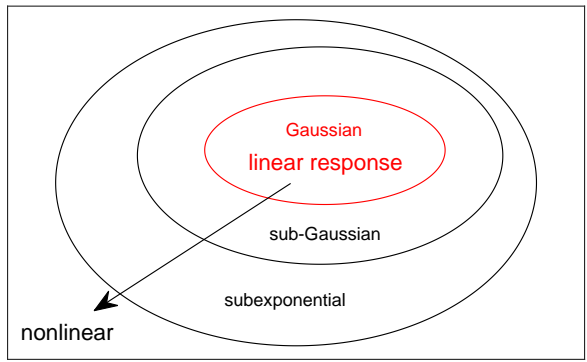,width=\linewidth}\caption{(Color online) A hierarchical structure of distributions. Whereas Gaussian distributions lay the foundation for linear response theory, we argue sub-Gaussian and subexponential distributions are relevant in general nonequilibrium thermodynamical states or processes where nonlinear response theory applies.} \label{fig:distribution_family}
  \end{center}
\end{figure}

Sub-Gaussian and subexponential distributions are important in modern statistics and machine learning \cite{Book:Martin,Book:Roman}, but they seem to be less known to the physics community. Roughly speaking, sub-Gaussian distributions are those possessing a tail that falls under a curve which is the tail of some Gaussian distribution lifted up, and subexponential distributions are those possessing a tail dominated by the uplifted tail of some exponential distribution. Gaussian distributions belong to the sub-Gaussian class, and sub-Gaussian distributions belong to the subexponential class. Such a hierarchical structure is depicted in Fig. \ref{fig:distribution_family}. We argue that beyond the linear response regime that corresponds to Gaussian distributions, in the nonlinear regime, sub-Gaussian and subexponential distributions are ubiquitous, as previous study shows. As an illustrative example, several fluctuation relations have the form $P(x)/P(-x)=e^x$, where $x$ represents some measure of irreversibility, its precise meaning depending on context \cite{RPP:Seifert}. Note that for all $x\geq0$ we have $P(X\leq -x)\leq e^{-x}$ \cite{note}. Hence the random variable $X$ has a tail that decays at least in an exponential way, thus $X$ is at least one-sided subexponential. Another physically relevant fact is that all bounded distributions are sub-Gaussian, hence one expects that sub-Gaussian distributions may be suitable for physical systems with finite states.

Rather than to analyze in detail the dynamics of a specific system, our aim in this work is to study how one can take advantage of the general properties of sub-Gaussian and subexponential distributions to study the nonlinear response theory in a unified way for different systems. However, the price we pay for this universality is that oftentimes we can only obtain results in the form of inequalities rather than equalities. This is reminiscent of the fact that the second law of thermodynamics is universally true for macroscopic systems, but the lower bound it provides on the entropy increase in a thermodynamical process can be substantially improved when more detailed information of the system in question is gained. The trade-off between universality and tightness of a bound is inevitable for thermodynamical theories involving inequalities.

To study the fluctuation-response relation out of equilibrium, we mainly follow the idea proposed in a recent work by Dechant and Sasa \cite{arXiv:Sasa}. By using the so-called sub-Gaussian (or subexponential) norm, we are able to further refine their results and provide a neat upper bound for the difference between expected values of an arbitrary sub-Gaussian (or subexponential) variable with respect to two distributions. Under different situations, the two distributions may represent the Boltzmann distributions of two equilibrium states connected by a not necessarily small perturbation, or they may correspond to the forward and backward processes in the setting of stochastic thermodynamics. In the former case, we also provide a bound for the entropy difference between two states, which turns out to be related to the energy fluctuation; while in the latter case, the bound we obtain is actually a generalized version of the thermodynamic uncertainty relation \cite{PRL:TUR01,PRL:TUR02}. If no other information is present, our result is a universal one in the sub-Gaussian or subexponential regime, respectively. Operational issues concerning these bounds for empirical data are then discussed. Nonasymptotic error bounds by using the sample mean instead of the expected value in our inequalities are derived, and we briefly address the way to estimate these norms.

In the following, we introduce the concepts and basic properties of sub-Gaussian and subexponential distributions, as well as their relevance in statistical physics, in Sec. \ref{sec:Sub-Gaussian and subexponential random variables}. Main theoretical results are established in Sec. \ref{sec:Concentration inequalities}. Using our results in an experimental setting is discussed in Sec. \ref{sec:Empirical}, and we conclude in Sec. \ref{sec:Conclusion}.

\section{Sub-Gaussian and subexponential random variables: Concepts and physical relevance}\label{sec:Sub-Gaussian and subexponential random variables}
Sub-Gaussian or subexponential random variables are defined based on the so-called sub-Gaussian or subexponential property. Such properties can be expressed in several different but equivalent ways \cite{Book:Martin,Book:Roman}. In this work, without loss of generality, we focus on centered random variables, which have zero mean. (Any random variable with a finite mean can always be transformed into a zero-mean random variable by subtracting its mean.) Physically, that means we are interested in the relative value of a stochastic quantity with respect to its mean at some reference state. Also note the mathematical fact that the qualitative sub-Gaussian or subexponential property will not be changed if a constant is subtracted from a random variable.

Next, we will briefly introduce the concepts of sub-Gaussian and subexponential random variables, respectively. More of their properties, and in particular, the concentration inequalities that are useful in an operational setting, are provided in the Appendices. Much of such information can be found in Refs. \cite{Book:Martin,Book:Roman}, but in this work it is organized in a way that we feel more suitable for application in statistical physics. Recall that our aim is to provide a universal approach to nonlinear response in different nonequilibrium regimes characterized by a hierarchy of distributions, Gaussian $\subset$ sub-Gaussian $\subset$ subexponential.

\subsection{Sub-Gaussian random variables}
A centered random variable $X$ is sub-Gaussian if for some $\sigma>0$ its moment generating function $\mathbb{E}e^{sX}$ satisfies
\begin{eqnarray}\label{eq:subG}
\mathbb{E}e^{sX}\leq e^{s^2\sigma^2/2}, \forall s\in\mathbb{R}.
\end{eqnarray}
Apparently, a Gaussian random variable is also sub-Gaussian. For our purposes, we also define the sub-Gaussian norm of $X$ as the infimum of $\sigma$ such that the sub-Gaussian property (\ref{eq:subG}) holds \cite{note:subGnorm}:
\begin{eqnarray}\label{eq:subGnorm}
\lVert X \rVert_\mathrm{G} = \inf\{ \sigma>0: \mathbb{E}e^{sX}\leq e^{s^2\sigma^2/2}, \forall s \in \mathbb{R} \}.
\end{eqnarray}
$\lVert X \rVert_\mathrm{G}^2$ is sometimes referred to as the optimal proxy variance. There are other ways to define a norm for sub-Gaussian variables, such as the Orlicz $\psi_2$-norm. These norms are strongly related and equivalent to each other up to a numerical constant factor; they emphasize on different aspects of the sub-Gaussian property. The reason why we choose $\lVert X \rVert_{\mathrm{G}}$ defined above as the sub-Gaussian norm is that $\lVert X \rVert_{\mathrm{G}}^2$ naturally reduces to the variance $\mathrm{var}(X)$ if $X$ is Gaussian. In general cases, we have
\begin{eqnarray}
\lVert X\rVert_\mathrm{G}^2\geq \mathrm{var}(X).
\end{eqnarray}
Non-trivial universal results could be obtained for qualitatively similar physical processes, in terms of the corresponding sub-Gaussian norm.

It is worth noting that when we speak of a sub-Gaussian distribution, it does not necessarily mean we refer to a family of distributions with a fixed parametric form like Gaussian distributions, which are parametrized by mean and variance. We can work with a sub-Gaussian variable as long as the condition (\ref{eq:subG}) holds, even if the explicit form of the distribution is unknown or intractable. The sub-Gaussian class of distributions is probably the simplest generalization of Gaussian distributions that can be relevant in nonequilibrium statistical physics. Typical distributions that are sub-Gaussian include Gaussian, Bernoulli, and in fact all bounded distributions. It is guaranteed that if $X\in[a,b]$, then
\begin{eqnarray}
\lVert X \rVert_\mathrm{G}\leq\frac{b-a}{2}.
\end{eqnarray}
This property is particularly relevant to finite-state Markovian dynamics.

More sub-Gaussian properties, especially the concentration inequality that is important in analyzing empirical data, are given in Appendix \ref{appendix:subgaussian}.

\subsection{Subexponential random variables}
Similarly, we briefly introduce the basic concept of (centered) subexponential variables here. Although there is no consensus on, and there exist different versions of, the definition of subexponential variables, these definitions are consistent with each other, all leading to the same kind of probability inequalities. In this work, a centered subexponential variable is a zero-mean random variable that satisfies
\begin{eqnarray}\label{eq:subE}
\mathbb{E}e^{sX}\leq e^{\sigma^2 s^2/2},\ \text{for}\ |s|\leq\frac{c_\mathrm{E}}{\sigma},
\end{eqnarray}
where $\sigma>0$ and $c_\mathrm{E}=(\sqrt{3}+1)/2$ is picked for later convenience. Note that, different than in (\ref{eq:subG}), here the range of $s$ is confined, thus apparently if $X$ is sub-Gaussian then it is simultaneously subexponential. We define the subexponential norm as \cite{note:subEnorm}
\begin{eqnarray}\label{eq:subEnorm}
\lVert X \rVert_\mathrm{E} = \inf\{\sigma>0: \mathbb{E}e^{sX}\leq e^{\sigma^2 s^2/2}, \text{for}\ |s|\leq\frac{c_\mathrm{E}}{\sigma}\}.
\end{eqnarray}
There are other possible norms for subexponential variables, such as the Orlicz $\psi_1$-norm, which are all strongly related and equivalent to each other up to a numerical constant factor. One nontrivial example of a centered subexponential variable is $X=Z^2-1$, where $Z\sim\mathcal{N}(0,1)$ is the standard Gaussian random variable. $Z^2$ follows the chi-square distribution with 1 degree of freedom, whose tail decays essentially in an exponential way, and $\mathbb{E}e^{sX}=\mathbb{E}e^{s(Z^2-1)}=\mathbb{E}e^{sZ^2}e^{-s}=e^{-s}/\sqrt{1-2s}$, which is only well defined for $s<1/2$. Numerically one can check that $\mathbb{E}e^{sX}\leq e^{3^2s^2/2}$ for $|s|\leq c_\mathrm{E}/3$. Hence $X$ is subexponential.

Our definition of subexponential variables (\ref{eq:subE}) is chosen in a way that not only ensures $\lVert X\rVert_\mathrm{E}^2=\lVert X\rVert_\mathrm{G}^2=\mathrm{var}(X)$ when $X$ is centered Gaussian, but also assumes a neatest form $e^{-t/\lVert X \rVert_\mathrm{E}}$ for the exponentially decaying tail without additional coefficients, as shown in Appendix \ref{appendix:subexponential}. Moreover, when a subexponential variable is not too distant away from its mean, i.e., $|X - \mathbb{E}X|\lesssim c_\mathrm{E}\lVert X \rVert_\mathrm{E}$, then it looks as if it is sub-Gaussian. Only the tail probability of a subexponential variable distinguishes it from a sub-Gaussian one.

\subsection{Statistical-physical relevance}\label{subsec:work_example}
To show that sub-Gaussian and subexponential random variables are relevant to statistical physics, let us consider an example here. In Ref. \cite{PRE:CJ}, Crooks and Jarzynski show that for a $d$-dimensional $n$-particle classical gas which is initially at inverse temperature $\beta$ (the Boltzmann constant $k_\mathrm{B}$ is set to 1 throughout this work) and undergoing an adiabatic and quasistatic compression process, the work performed to the system follows a gamma (rather than Gaussian) distribution. For simplicity, let us consider the case in two dimensions where the gas is compressed to half its initial volume; then the associated work distribution $P(W)$ is
\begin{eqnarray}
P(W) = \frac{\beta^n}{\Gamma(n)}W^{n-1}e^{-\beta W}\ (W>0).\notag
\end{eqnarray}
Based on the properties of gamma distributions, we know the mean work is $\mathbb{E}W=n/\beta$, the variance is $\mathrm{var}(W)=n/\beta^2$, and the moment generating function is $\mathbb{E}e^{sW}=(1-s/\beta)^{-n}$ with $s/\beta < 1$. Hence the moment generating function of the centered work $\Delta W \equiv W-\mathbb{E}W$ is
\begin{eqnarray}\label{eq:centeredW}
\mathbb{E}e^{s(W-\mathbb{E}W)} = e^{-sn/\beta}(1-s/\beta)^{-n}\ (s/\beta<1).
\end{eqnarray}
After some algebra as given in Appendix \ref{appendix:subE_norm}, we know
\begin{eqnarray}
\lVert\Delta W\rVert_\mathrm{E} \leq \sigma_\mathrm{E}\equiv\sqrt{\mathrm{var}(W)} + \frac{c_\mathrm{E}}{\beta}.\notag
\end{eqnarray}

\begin{figure}[tbp]
  \begin{center}
  \epsfig{figure=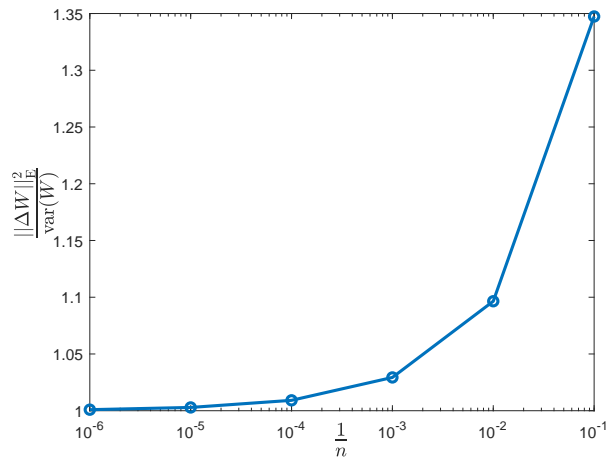,width=\linewidth}\caption{(Color online) In the example of the compression process of a classical gas, the distribution of work performed is subexponential. However, as the number of gas particles $n$ increases, the distribution can be well approximated by a Gaussian, as evidenced by the fact that the squared subexponential norm of centered work decreases to the variance of work.} \label{fig:subEvsN}
  \end{center}
\end{figure}

One can also numerically check that $\Delta W$ is subexponential and calculate the corresponding subexponential norm based on its definition. (It is useful to set the initial value of the trial subexponential norm as $\mathrm{var}(W)$ to start the search.) Concretely, setting $\beta=1$, we in Fig. \ref{fig:subEvsN} plot how the ratio $\lVert\Delta W\rVert_\mathrm{E}^2/\mathrm{var}(W)$ changes as $n$ is increased. One can see when $n\to\infty$, $\lVert\Delta W\rVert_\mathrm{E}^2/\mathrm{var}(W)\to 1$, indicating that in this limit, the work distribution can be well approximated by a Gaussian. In fact, depending on the requirement of and/or the approximation made to the distribution, there exists a hierarchy of results that are thermodynamically relevant.

(i) Distribution-free universal result, which is obtained essentially with no requirement. By the Jarzynski equality \cite{PRL:Jarzynski}, we have $\mathbb{E}e^{-\beta W}=e^{-\beta\Delta F}$, where $\Delta F$ is the change in Helmholtz free energy, whose precise meaning is given in Ref. \cite{PRE:CJ}. Thus letting $s=-\beta$ in Eq. (\ref{eq:centeredW}), and noting the convexity of the exponential function, one readily recovers that
\begin{eqnarray}\label{eq:2ndlawresult}
\mathbb{E}W\geq \Delta F,
\end{eqnarray}
without further knowing $W$ is subexponential.

(ii) Gaussian approximation. For $s/\beta\in[-1,1)$, Taylor expansion gives
\begin{eqnarray}
\ln\mathbb{E}e^{s(W-\mathbb{E}W)} = \frac{s^2}{\beta^2}\frac{n}{2}+\frac{s^3}{\beta^3}\frac{n}{3}+\ldots.\notag
\end{eqnarray}
To the leading term in $s$, and note $n/\beta^2=\mathrm{var}(W)$, one finds $\Delta W$ approximately follows a Gaussian distribution $\mathcal{N}(0,n/\beta^2)$. Now, if one lets $s=-\beta$, then also by the Jarzynski equality, one obtains
\begin{eqnarray}\label{eq:linearresponseresult}
\mathbb{E}W= \Delta F + \frac{1}{2}\beta\mathrm{var}(W).
\end{eqnarray}
This formula is often seen as a linear response result. However, it is not exact, since when setting $s=-\beta$ apparently the higher order terms of $s$ cannot be simply discarded. It is a bit subtle that Gaussian approximation is good when $|s|$ is small, but in this example $\lvert -\beta\rvert$ is always not small enough to make this approximation accurate enough. Nonetheless, compared with (i), the Gaussian approximation helps one gain a more informative (though inaccurate) result.

(iii) Sub-Gaussian approximation. By the concentration inequality of subexponential variables given in Appendix \ref{appendix:subexponential}, this approximation implies that $|\Delta W|\lesssim c_\mathrm{E}\lVert\Delta W\rVert_\mathrm{E}$ is predominantly common. That is to say, the deviation of $W$ from its mean $\mathbb{E}W$ rarely reaches the ``boundary'' $c_\mathrm{E}\lVert\Delta W\rVert_\mathrm{E}$, and $\Delta W$ is essentially a sub-Gaussian variable with norm $\lVert\Delta W\rVert_\mathrm{G}\approx \lVert\Delta W\rVert_\mathrm{E}$. By the sub-Gaussian property, we can approximately obtain $-\beta(\Delta F -\mathbb{E}W)=\ln\mathbb{E}e^{-\beta(W-\mathbb{E}W)}\leq \lVert\Delta W\rVert_\mathrm{G}^2\beta^2/2$, hence
\begin{eqnarray}
\mathbb{E}W\leq \Delta F + \frac{1}{2}\beta\lVert\Delta W\rVert_\mathrm{G}^2.
\end{eqnarray}
Note that $\lVert\Delta W\rVert_\mathrm{G}^2\geq\mathrm{var}(W)=n/\beta^2$, hence physically bigger $n$ and smaller $\beta$ make the boundary $c_\mathrm{E}\lVert\Delta W\rVert_\mathrm{G}$ more difficult to reach, resulting in a better sub-Gaussian approximation.

In general, if $\Delta W$ is strictly sub-Gaussian (for example, there is restriction on the upper bound of $|W|$ for some physical process), then in this case we can have an exact result that
\begin{eqnarray}
\Delta F \leq \mathbb{E}W \leq \Delta F + \frac{1}{2}\beta \lVert\Delta W\rVert_\mathrm{G}^2.
\end{eqnarray}

(iv) Subexponential case without any approximation. This is the most general situation, and for this specific example we have exactly that
\begin{eqnarray}\label{eq:EW_upper_bound}
\mathbb{E}W\leq \Delta F + \frac{1}{2}\beta \sigma_\mathrm{E}^2,
\end{eqnarray}
with $\sigma_\mathrm{E}$ defined above. Note that, even though $\sigma_\mathrm{E}$ is defined based on $|s|\leq c_\mathrm{E}/\sigma_\mathrm{E} <\beta$, the property of the moment generating function in this model assures that the above inequality holds for $s=-\beta$; see Appendix \ref{appendix:subE_norm} for details.

In general, let $f(s)\equiv\ln\mathbb{E}e^{s(W-\mathbb{E}W)}$; then $f(0)=f'(0)=0$ and $f''(s)>0$ as long as $f(s)$ is well defined \cite{note:f}. Following similar steps in Appendix \ref{appendix:subE_norm}, especially (\ref{eq:fss_upper_bound}), one can see if, for all $s\in[-\beta,0]$, $f''(s)$ is bounded from above by a constant, also denoted as $\sigma_\mathrm{E}^2$, then $\mathbb{E}W$ can be upper bounded as in (\ref{eq:EW_upper_bound}). Under mild conditions, such a bound always exists. For example, we can take $\sigma_\mathrm{E}^2\equiv\max_{s\in[-\beta,0]}f''(s)$; however, the physical meaning of such a bound is not fully clear at present. On the other hand, if the range of $s$ to calculate the subexponential norm also includes $s=-\beta$ in some physical process, then we can have an exact range for $\mathbb{E}W$:
\begin{eqnarray}
\Delta F \leq \mathbb{E}W \leq \Delta F + \frac{1}{2}\beta \lVert\Delta W\rVert_\mathrm{E}^2.
\end{eqnarray}

From this example, we can see both sub-Gaussian and subexponential distributions could be relevant in statistical physics. By taking advantage of their properties, one might be able to go beyond the results from the second law of thermodynamics and linear response theory, obtaining a system-dependent upper bound of $\mathbb{E}W$.

\section{Application in nonlinear response}\label{sec:Concentration inequalities}
\subsection{General theory}\label{sec:Response}
Now we investigate how the introduction of sub-Gaussian and subexponential variables can provide useful information in nonlinear response. We mainly follow Dechant and Sasa's idea \cite{arXiv:Sasa}. In the general setting, we have a physical system, characterized by a probability distribution $P\in\mathcal{P}$, where $\mathcal{P}$ is a probability family on a measurable space $\Omega$. We assume for simplicity here that, for any $P\in\mathcal{P}$, $P$ has a density function with respect to some dominating measure. Abusing the notation a bit, we may also denote the density function as $P$. Such distributions can be those that characterize steady states or stochastic trajectories. For a random variable $X$, we denote its centered version as $\Delta X=X - \mathbb{E}X$. First let us use $P_0$ as the reference probability that describes the unperturbed state or the forward process, and $P_1$ as the distribution for the perturbed state or the backward process. One is interested how the change in distribution affects the ensemble average of $X$. Suppose for now $\mathbb{E}_0X-\mathbb{E}_1X\geq 0$; then starting from the moment generating function of $\Delta X$ with respect to $P_1$ (when well defined), we have for $s\geq0$ that
\begin{eqnarray}
\ln\mathbb{E}_1e^{s\Delta X} &=& \ln \left( \int_\Omega e^{s\Delta X(\omega)} P_1(\omega) d\omega\right)\notag\\
&=& \ln \left( \int_\Omega e^{s[X(\omega)-\mathbb{E}_1X]} \frac{P_1(\omega)}{P_0(\omega)} P_0(\omega) d\omega\right)\notag\\
&=&\ln \mathbb{E}_0 \left( e^{s[X-\mathbb{E}_1X]} \frac{P_1}{P_0} \right)\notag\\
&\geq& \mathbb{E}_0\ln\left( e^{s[X-\mathbb{E}_1X]} \frac{P_1}{P_0} \right)\ (\text{by\ Jensen's})\notag\\
&=& s(\mathbb{E}_0X-\mathbb{E}_1X) - D_\mathrm{KL}(P_0\rVert P_1),\notag
\end{eqnarray}
where $D_\mathrm{KL}(P_0\rVert P_1) = -\int_\Omega \ln(P_1/P_0)P_0d\omega$ is defined to be the Kullback-Leibler divergence between $P_0$ and $P_1$. It is always nonnegative and equal to 0 only when $P_0=P_1$. Rearrange to obtain
\begin{eqnarray}
\mathbb{E}_0X - \mathbb{E}_1X \leq \frac{1}{s}\left[\ln\mathbb{E}_1e^{s\Delta X} + D_\mathrm{KL}(P_0\rVert P_1)\right]\notag
\end{eqnarray}
for all $s\geq0$ when $\mathbb{E}_1e^{s\Delta X}$ is well defined. Similarly, if $\mathbb{E}_0X-\mathbb{E}_1X<0$, then for $s\geq0$ we have
\begin{eqnarray}
\ln\mathbb{E}_1e^{-s\Delta X} \geq -s(\mathbb{E}_0X-\mathbb{E}_1X) - D_\mathrm{KL}(P_0\rVert P_1),\notag
\end{eqnarray}
which implies
\begin{eqnarray}
\mathbb{E}_1X - \mathbb{E}_0X \leq \frac{1}{s}\left[\ln\mathbb{E}_1e^{-s\Delta X} + D_\mathrm{KL}(P_0\rVert P_1)\right]\notag
\end{eqnarray}
for all $s\geq0$ when $\mathbb{E}_1e^{-s\Delta X}$ is well defined. Hence, combining these results, we have a Chernoff-like inequality that
\begin{eqnarray}\label{eq:thermo_chernoff}
|\mathbb{E}_1X - \mathbb{E}_0X| \leq \inf_{s\geq 0} \left\{ \frac{1}{s}\left[\ln\mathbb{E}_1e^{\xi s\Delta X} + D_\mathrm{KL}(P_0\rVert P_1)\right] \right\},\notag\\
\end{eqnarray}
where $\xi=\text{sgn}(\mathbb{E}_0X - \mathbb{E}_1X)$. This is one central result in Ref. \cite{arXiv:Sasa}.

On the other hand, due to symmetry, it is straightforward to have that
\begin{eqnarray}\label{eq:thermo_chernoff_2}
|\mathbb{E}_1X - \mathbb{E}_0X| \leq \inf_{s\geq 0} \left\{ \frac{1}{s}\left[\ln\mathbb{E}_0e^{\xi s\Delta X} + D_\mathrm{KL}(P_1\rVert P_0)\right] \right\}.\notag\\
\end{eqnarray}

Hence, one can take the minimum of these two upper bounds as $b=\min\{b_1, b_2\}$, where $b_1$ and $b_2$ denote the above bounds in (\ref{eq:thermo_chernoff}) and (\ref{eq:thermo_chernoff_2}). Note that neither of $b_1$ and $b_2$ is symmetric in $P_0$ and $P_1$. One might tend to construct symmetric bounds in the form $|\mathbb{E}_1X - \mathbb{E}_0X| \leq (b_1 + b_2)/2$, $|\mathbb{E}_1X - \mathbb{E}_0X| \leq \sqrt{b_1b_2}$, etc.; however, these bounds are less tight than $b$. In the following, without loss of generality, we will assume (\ref{eq:thermo_chernoff}) provides a tighter bound than (\ref{eq:thermo_chernoff_2}) does. But before we show how the bound could be explicitly expressed in terms of sub-Gaussian or subexponential norm, rather than in terms of cumulants as in Ref. \cite{arXiv:Sasa}, let us consider two important situations where $D_\mathrm{KL}$ can be (at least partially) expressed by thermodynamical quantities.

\subsubsection{$D_\mathrm{KL}$ between two equilibrium states}
First, let us consider the case that $P_0$ and $P_1$ are the corresponding Boltzmann distributions at two states with Hamiltonians $H_0$ and $H_1$, respectively. Hence $P_0=e^{-\beta H_0}/Z_0$ and $P_1=e^{-\beta H_1}/Z_1$, where $\beta$ denotes the inverse temperature of the system and $Z_{0,1}$ are partition functions. Then $D_\mathrm{KL}(P_0\rVert P_1)$ can be written as
\begin{eqnarray}
D_\mathrm{KL}(P_0\rVert P_1) &=& \mathbb{E}_0\ln\left(\frac{e^{-\beta H_0}/Z_0}{e^{-\beta H_1}/Z_1}\right)\notag\\
&=& -\beta\mathbb{E}_0(H_0-H_1) + \ln\left(\frac{Z_1}{Z_0}\right).\notag
\end{eqnarray}
Note $-\ln Z_{0,1}/\beta = F_{0,1}$ where $F_{0,1}$ are the Helmholtz free energies, and $\mathbb{E}_0H_0$ is the internal energy $U_0$, hence $D_\mathrm{KL}(P_0\rVert P_1) = -\beta(U_0-\mathbb{E}_0H_1) - \beta(F_1-F_0)$. The only term that has no direct thermodynamical correspondence is $\mathbb{E}_0H_1$, which can be written as $\mathbb{E}_0H_1 = \mathbb{E}_1H_1 + (\mathbb{E}_0H_1-\mathbb{E}_1H_1)
= U_1 + \int_\Omega H_1(P_0-P_1)d\omega$. By the thermodynamical relation $F=U-S/\beta$, where $S$ is the entropy, one finds that
\begin{eqnarray}\label{eq:DKL_integral_term}
D_\mathrm{KL}(P_0\rVert P_1) &=& S_1 - S_0 + \beta(\mathbb{E}_0H_1-\mathbb{E}_1H_1)\notag\\
&=& S_1 - S_0 + \beta\int_\Omega H_1(P_0-P_1)d\omega.
\end{eqnarray}
Interestingly, if $H_1$ is Lipschitz with constant $L_1$, then the integral on the right-hand side can be bounded in terms of the Wasserstein distance \cite{Book:Villani} as $\beta\int_\Omega H_1(P_0-P_1)d\omega \leq \beta L_1W_1(P_0,P_1)$, where $W_1$ is the 1-Wasserstein distance between $P_0$ and $P_1$ \cite{note:W1}. Thus, as a byproduct, we find that the entropy change can be bounded by
\begin{eqnarray}
S_1-S_0\geq D_\mathrm{KL}(P_0\rVert P_1) - \beta L_1W_1(P_0,P_1).
\end{eqnarray}
When $P_0=P_1$, the bound is tight. However, without more information, there is no guarantee on the tightness of this bound for general $P_0$ and $P_1$. Recently there have been some works to address the thermodynamical relevance of Wasserstein distance \cite{arXiv:Dechant,arXiv:TVV}; however, in this work we will come back to Eq. (\ref{eq:DKL_integral_term}) later using sub-Gaussian or subexponential norm.

By the way, it is worth noting that the explicit form of $D_\mathrm{KL}$ can be substantially simplified in the linear response regime. If $H_1 = H_0 + \varepsilon A$, where $\varepsilon$ is a small parameter, then as shown in Appendix \ref{appendix:KL_linear_response}, we have that
\begin{eqnarray}\label{eq:DKL_linear_response}
D_\mathrm{KL}(P_0\rVert P_1)\approx D_\mathrm{KL}(P_1\rVert P_0)\approx\frac{1}{2}\varepsilon^2\beta^2\mathrm{var}_0A.
\end{eqnarray}

\subsubsection{$D_\mathrm{KL}$ between forward and backward processes}
Second, as is well known, if $P_0$ denotes the probability for a forward path $\omega$, and $P_1$ is the backward probability for the time-reversed path $\omega^\dagger$, then $D_\mathrm{KL}(P_0\rVert P_1)$ can be interpreted as the total entropy production $\Delta S$ \cite{RPP:Seifert}:
\begin{eqnarray}\label{eq:path}
D_\mathrm{KL}(P_0\rVert P_1) = \Delta S.
\end{eqnarray}
This relation may be obtained in different settings, either in the case of Langevin dynamics with a well selected probability density function at the final state \cite{PRL:Seifert2005}, or in the case of a Hamiltonian system controlled by an external protocol first and then reconnected to a heat bath \cite{PRL:VdB2007}.

In the following, we will use the sub-Gaussian or subexponential property to deal with the first term $\ln\mathbb{E}_1e^{\xi s\Delta X}$ in the upper bound (\ref{eq:thermo_chernoff}).

\subsection{Sub-Gaussian regime}
If $P_0$ and $P_1$ are sub-Gaussian, then by (\ref{eq:subG}) and (\ref{eq:subGnorm}), we have
\begin{eqnarray}
\ln\mathbb{E}_1e^{\xi s\Delta X} \leq \frac{1}{2} \lVert \Delta X \rVert_\mathrm{1G}^2 (\xi s)^2 = \frac{1}{2} \lVert \Delta X \rVert_\mathrm{1G}^2 s^2,\notag
\end{eqnarray}
where $\lVert \Delta X \rVert_\mathrm{1G}$ is the sub-Gaussian norm of $\Delta X$ with respect to $P_1$. Inserting this into (\ref{eq:thermo_chernoff}), we have
\begin{eqnarray}\label{eq:subG_X_mean_diff}
|\mathbb{E}_1X - \mathbb{E}_0X| &\leq& \inf_{s\geq 0}\left\{ \frac{1}{2}\lVert \Delta X \rVert_\mathrm{1G}^2 s + \frac{1}{s}D_\mathrm{KL}(P_0\rVert P_1) \right\}\notag\\
&=& \lVert \Delta X \rVert_\mathrm{1G}\sqrt{2D_\mathrm{KL}(P_0\rVert P_1)}.
\end{eqnarray}
Note that for nonnegative $x$ and $y$, the inequality $x+y\geq 2\sqrt{xy}$ always holds with equality attained when $x=y$, hence the infimum is achieved when $D_\mathrm{KL}(P_0\rVert P_1)/s=\lVert \Delta X \rVert_\mathrm{1G}^2s/2$. Inequality (\ref{eq:subG_X_mean_diff}) provides a universal bound on the relative difference between means of $X$ in terms of the sub-Gaussian norm.

\begin{figure}[tbp]
  \begin{center}
  \epsfig{figure=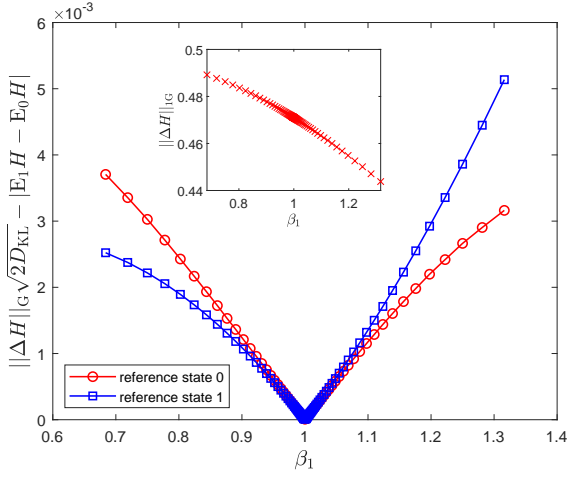,width=\linewidth}\caption{(Color online) For a two-level system at inverse temperatures $\beta_0$ and $\beta_1$, the absolute internal energy difference is bounded according to (\ref{eq:subG_X_mean_diff}), no matter whether the reference state for calculating $D_\mathrm{KL}$ corresponds to $\beta_0$ or $\beta_1$. We fix $\beta_0=1$ and change $\beta_1$ to obtain the results; the inset shows how $\lVert\Delta H\rVert_{1G}$ varies accordingly.} \label{fig:subG_eg}
  \end{center}
\end{figure}

\subsubsection{Numerical example of a two-level system}
As a concrete example, let us consider a simple one-particle two-level system with energy levels 0 and 1, and we want to bound the internal energy difference at two inverse temperature $\beta_0$ and $\beta_1$. In this case, $X=H$ is the Hamiltonian, and each term in (\ref{eq:subG_X_mean_diff}) can be numerically calculated. In particular, it is worth noting that $H$ is bounded in $[0,1]$ at both inverse temperatures, and $\lVert \Delta H \rVert_\mathrm{G}\leq 0.5$ is guaranteed \cite{Book:Martin,Book:Roman}. Setting $\beta_0=1$, in Fig. \ref{fig:subG_eg} we show that for various choices of $\beta_1$ the absolute difference $|\mathbb{E}_1H - \mathbb{E}_0H|$ is indeed bounded by $\lVert \Delta H \rVert_\mathrm{G}\sqrt{2D_\mathrm{KL}}$, no matter at which reference state $D_\mathrm{KL}$ is obtained. Reference state 0 refers to the state with inverse temperature $\beta_0$, and in this case $D_\mathrm{KL}(P_0\rVert P_1)$ is calculated. Similarly, $D_\mathrm{KL}(P_1\rVert P_0)$ is calculated at reference state 1. In both cases, $D_\mathrm{KL}$ can be easily obtained since the probability distributions of $\Delta H$ are discrete. Sub-Gaussian norms are calculated based on the definition. (It is helpful to set the initial value of them as 0.5 in the searching algorithm.) The inset of Fig. \ref{fig:subG_eg} shows how $\lVert\Delta H\rVert_{1G}$ varies as $\beta_1$ is changed, which is always less than 0.5 as expected.

\subsubsection{Bound for entropy change between two equilibrium states}
Now let us consider Eq. (\ref{eq:DKL_integral_term}) again. If $\Delta H_1$ is sub-Gaussian, which could be possible for a system with bounded energy, then insert Eq. (\ref{eq:DKL_integral_term}) into (\ref{eq:subG_X_mean_diff}). Taking $X=H_1$, we have
\begin{eqnarray}
&&|\mathbb{E}_1H_1 - \mathbb{E}_0H_1|\notag\\
&\leq& \lVert \Delta H_1 \rVert_\mathrm{1G}\sqrt{2D_\mathrm{KL}(P_0\rVert P_1)}\notag\\
&=& \lVert \Delta H_1 \rVert_\mathrm{1G}\sqrt{2[S_1 - S_0 + \beta(\mathbb{E}_0H_1-\mathbb{E}_1H_1)]}.\notag
\end{eqnarray}
Solving this inequality, we get bounds for $\mathbb{E}_0H_1 - \mathbb{E}_1H_1$ as:
\begin{eqnarray}
&&\ \beta\lVert \Delta H_1 \rVert_\mathrm{1G}^2\left(1-\sqrt{1+\frac{2(S_1-S_0)}{\lVert \Delta H_1 \rVert_\mathrm{1G}^2\beta^2}}\right)\notag\\ \leq &&\ \mathbb{E}_0H_1 - \mathbb{E}_1H_1\notag\\ \leq &&\ \beta\lVert \Delta H_1 \rVert_\mathrm{1G}^2\left(1+\sqrt{1+\frac{2(S_1-S_0)}{\lVert \Delta H_1 \rVert_\mathrm{1G}^2\beta^2}}\right),\notag
\end{eqnarray}
which holds under the condition that
\begin{eqnarray}\label{eq:entropy_change_condition}
S_1-S_0\geq -\frac{1}{2}\beta^2\lVert \Delta H_1 \rVert_\mathrm{1G}^2.
\end{eqnarray}

There are several thermodynamical implications of this result. First, we establish a connection between the difference in the ensemble averages of a Hamiltonian and the entropy difference at two states. Second, although (\ref{eq:entropy_change_condition}) is a mathematical requirement, physically we know that this must hold, and this implies that there is a bound on the entropy change between two states, which is given by the property $\lVert \Delta H_1 \rVert_\mathrm{1G}$. At the first sight, it does not seem to make sense since there is no information of the other state (state ``0'') involved, however, note that we have made an assumption that it is (\ref{eq:thermo_chernoff}) rather than (\ref{eq:thermo_chernoff_2}) that gives the tighter bound. Hence the information of state ``0'' is used. Actually, (\ref{eq:entropy_change_condition}) has a natural twin by switching the indices 0 and 1:
\begin{eqnarray}\label{eq:entropy_change_condition_02}
S_0-S_1\geq -\frac{1}{2}\beta^2\lVert \Delta H_0 \rVert_\mathrm{0G}^2.
\end{eqnarray}
If $S_1>S_0$, then (\ref{eq:entropy_change_condition}) is trivial, but (\ref{eq:entropy_change_condition_02}) upper bounds $S_1-S_0$; while if $S_0>S_1$, then (\ref{eq:entropy_change_condition_02}) is trivial, but (\ref{eq:entropy_change_condition}) upper bounds $S_0-S_1$. Hence we can summarize that
\begin{eqnarray}\label{eq:entropy_change_condition_03}
|S_1-S_0|\leq\max\left\{\frac{1}{2}\beta^2\lVert \Delta H_0 \rVert_\mathrm{0G}^2, \frac{1}{2}\beta^2\lVert \Delta H_1 \rVert_\mathrm{1G}^2\right\}.\notag\\
\end{eqnarray}
This result is a most general one concerning entropy change in the sub-Gaussian regime.

One can verify that (\ref{eq:entropy_change_condition_03}) holds in the simpler linear response case where $H_1=H_0+\varepsilon A$. By the thermodynamical relation $S=-\beta\partial\ln Z/\partial\beta - \ln Z$, as well as the results in Appendix \ref{appendix:KL_linear_response}, we can find $S_1-S_0=-\beta\partial\ln(Z_1/Z_0)/\partial\beta-\ln(Z_1/Z_0)\approx 2\varepsilon\beta\mathbb{E}_0A\sim O(\varepsilon^1)$, while $\lVert \Delta H\rVert_\mathrm{G}^2$ becomes close to $\mathrm{var}(\Delta H)=\partial^2\ln Z/\partial\beta^2$, and in any state it does not depend on $A$ to the order of $O(\varepsilon^0)$. Hence (\ref{eq:entropy_change_condition_03}) naturally holds. As the perturbation $\varepsilon$ becomes larger, our result (\ref{eq:entropy_change_condition_03}) is expected to be nontrivial.

For some other sub-Gaussian physical quantity $X$ in the linear response regime, and in particular for $X$ that can be well approximated by a Gaussian, we have $\lVert \Delta X \rVert_\mathrm{1G}^2 \approx \mathrm{var}_1(\Delta X)$. As also shown in Appendix \ref{appendix:KL_linear_response}, $|\mathbb{E}_1X-\mathbb{E}_0X|\approx|\varepsilon\beta\mathrm{cov}_0(X,A)|$, and, to the leading order, $\mathrm{var}_1(\Delta X) = \mathrm{var}_0(\Delta X)$. Hence, by Eqs. (\ref{eq:DKL_linear_response}) and (\ref{eq:subG_X_mean_diff}) we obtain
\begin{eqnarray}
|\mathrm{cov}_0(X,A)| &\leq& \sqrt{\mathrm{var}_0(\Delta X)}\sqrt{\mathrm{var}_0A}\notag\\
&=& \sqrt{\mathrm{var}_0X}\sqrt{\mathrm{var}_0A},\notag
\end{eqnarray}
which shows that (\ref{eq:subG_X_mean_diff}) coincides with the Cauchy-Schwarz inequality in linear response.

\subsubsection{Sub-Gaussian thermodynamic uncertainty
relation}\label{subsec:subG_norm_bound}
While for the path thermodynamics example (\ref{eq:path}), we have
\begin{eqnarray}
|\mathbb{E}_1X - \mathbb{E}_0X|\leq\lVert \Delta X \rVert_{1\mathrm{G}}\sqrt{2\Delta S}.\notag
\end{eqnarray}
If $X$ changes sign under time reversal, then we further have
\begin{eqnarray}\label{eq:subG_thermo_inequality}
2(\mathbb{E}X)^2 \leq \lVert \Delta X \rVert_{\mathrm{G}}^2\Delta S.
\end{eqnarray}
Note there is no need to specify with respect to which distribution the mean and sub-Gaussian norm are taken due to the time reversal operation. Again, when the Gaussian approximation is valid, we formally recover the thermodynamic uncertainty relation that \cite{PRL:TUR01,PRL:TUR02}
\begin{eqnarray}\label{eq:TUR}
2(\mathbb{E}X)^2 \leq \mathrm{var}(X)\Delta S.
\end{eqnarray}
It is worth noting, however, that Gaussian distributions are sufficient but not necessary for this relation to hold. It can be valid in more general settings. For example, it is known that it holds for continuous-time Markov dynamics. Nonetheless, it may also fail for discrete-time dynamics as shown below. Our theoretical results (\ref{eq:subG_X_mean_diff}) and (\ref{eq:subG_thermo_inequality}) hold in the absence of information about the detailed dynamics, thus substantially pushing existing bounds to the sub-Gaussian nonlinear regime, at the expense of bound tightness in some cases due to the fact that $\lVert \Delta X \rVert_{\mathrm{G}}^2$ is always no less than $\mathrm{var}(X)$.

Recently, another bound has been derived for general dynamics based on the fluctuation theorem \cite{PRL:TURFT}, which states that
\begin{eqnarray}\label{eq:TURFT}
2(\mathbb{E}X)^2 \leq \mathrm{var}(X)(e^{\Delta S}-1).
\end{eqnarray}
Since it is always true that $\lVert \Delta X \rVert_{\mathrm{G}}^2\geq \mathrm{var}(X)$ but $\Delta S \leq e^{\Delta S}-1$, there is no definite conclusion whether our bound (\ref{eq:subG_thermo_inequality}) or the fluctuation theorem based bound (\ref{eq:TURFT}) is tighter in general. Nonetheless, as discussed in Ref. \cite{arXiv:Sasa}, the $e^{\Delta S}$ term in (\ref{eq:TURFT}) might lead to a loose bound in the large-system long-time case, while using the sub-Gaussian norm based bound (\ref{eq:subG_thermo_inequality}) may be more favorable.

In Ref. \cite{arXiv:Sasa}, a finite-state discrete-time Markov model is proposed to show the possible failure of the thermodynamic uncertainty relation (\ref{eq:TUR}). We will analyze the same model here, and show that this model falls in the sub-Gaussian category, where results in this work apply.

A random walker hops on a ring of $N$ sites at times $t=1,\ldots,M$. There are two relevant probabilities $q$ and $p$: $q$ stands for the probability of leaving a site at some $t$, and $p$ is the probability of hopping in the positive direction. The distance between two nearest sites is $1/N$. The quantity of interest is $X=r$, which is the displacement right after time $M$. For this model at the steady state, one can calculate exactly that
\begin{eqnarray}
&&\mathbb{E}r = \frac{M}{N}q(2p-1),\notag\\
&&\mathrm{var}(r) = \frac{M}{N^2}q\left[1-q(2p-1)^2\right],\notag\\
&&\Delta S = M\left[
                    qp\ln\left(\frac{p}{1-p}\right)
                  + q(1-p)\ln\left(\frac{1-p}{p}\right)
                    \right].\notag
\end{eqnarray}

\begin{figure}[tbp]
  \begin{center}
  \epsfig{figure=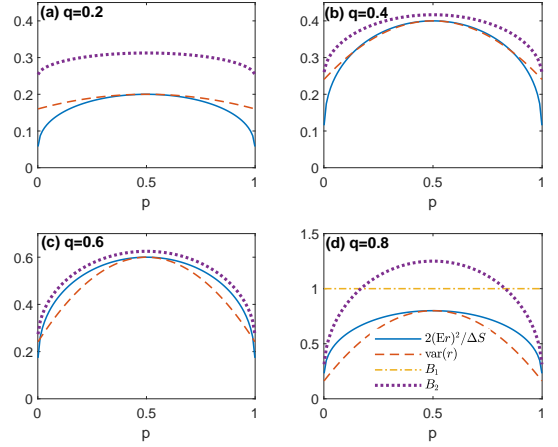,width=\linewidth}\caption{(Color online) Up to a factor $M/N^2$, four quantities $2(\mathbb{E}r)^2/\Delta S$, $\mathrm{var}(r)$, $B_1$, and $B_2$ are plotted for a finite-state discrete-time Markov model (see text). $B_1$ and $B_2$ always upper bound $2(\mathbb{E}r)^2/\Delta S$, while $\mathrm{var}(r)$ will fail as an upper bound as $q$ is increased. Note $B_1=1$ is only plotted in (d), since in other cases it is not a very informative bound.} \label{fig:subG_norm_bound}
  \end{center}
\end{figure}

As long as $p$ and $q$ are given, it is possible to numerically calculate the corresponding norm. However, it is also possible to explicitly upper bound the norm for any choice of $p$ and $q$. This approach is particularly suitable when a physical insight is preferred.

First, observe that $r=\sum_{t=1}^M l_t$, where $l_t$ is obviously bounded in $[-1/N,1/N]$ and thus is sub-Gaussian. Hence we immediately have
\begin{eqnarray}
\lVert \Delta r \rVert_\mathrm{G} \leq \frac{\sqrt{M}}{N}\equiv \sqrt{B_1}.
\end{eqnarray}
But for this problem, it is also possible to have another more informative upper bound of $\lVert \Delta r \rVert_\mathrm{G}$, as detailed in Appendix \ref{appendix:subGnorm_bound}:
\begin{eqnarray}
\lVert \Delta r \rVert_\mathrm{G} \leq \frac{\sqrt{M}}{N}\sqrt{\frac{1}{4} + \frac{1}{2}\frac{q}{1-q}\sqrt{p(1-p)}}\equiv \sqrt{B_2}.
\end{eqnarray}

We can use $B_1$ and $B_2$ to bound the quantity $2(\mathbb{E}r)^2/\Delta S$, and compare the results with (\ref{eq:TUR}). Up to a common factor $M/N^2$, we plot $2(\mathbb{E}r)^2/\Delta S$, $\mathrm{var}(r)$, $B_1$, and $B_2$ in Fig. \ref{fig:subG_norm_bound}. One can find that as $q$ is increased, the variance $\mathrm{var}(r)$ fails to serve as an upper bound for $2(\mathbb{E}r)^2/\Delta S$, but $B_1$ and $B_2$ still work. $B_1$ is better for extremely large $q$, while $B_2$ is more suitable for moderate $q$. Hence, as $q$ is gradually increased from 0, $\mathrm{var}(r)$, $B_2$, and $B_1$ provide a proper (in terms of tightness) bound in turn, and the classical thermodynamic uncertainty relation (\ref{eq:TUR}) has to be replaced with the more general sub-Gaussian version (\ref{eq:subG_thermo_inequality}).

Although (\ref{eq:TUR}) can be obtained from (\ref{eq:subG_thermo_inequality}) when $X$ is Gaussian, it also holds when $X$ is non-Gaussian, as shown in the above example for small $q$. In the latter case, we might say that the distribution of $X$ is ``close to'' a Gaussian, and (\ref{eq:TUR}) is a stronger result than (\ref{eq:subG_thermo_inequality}) for this certain range of $q$. As argued in Ref. \cite{arXiv:Sasa}, this $q$-dependence issue is related to the nature of the dynamics being discrete rather than continuous. Nonetheless, the sub-Gaussian bound (\ref{eq:subG_thermo_inequality}) is universal without any assumption of the dynamics; it only depends on the fact that, at any given time, $X$ is sub-Gaussian. This observation reflects the tradeoff between universality and tightness of bounds.

\subsection{Subexponential regime}

In this case, one might directly insert (\ref{eq:subE}) and (\ref{eq:subEnorm}) into (\ref{eq:thermo_chernoff}) to get
\begin{eqnarray}\label{eq:subE_X_mean_diff_vertex}
|\mathbb{E}_1X - \mathbb{E}_0X| \leq \lVert \Delta X \rVert_\mathrm{1E}\sqrt{2D_\mathrm{KL}(P_0\rVert P_1)},
\end{eqnarray}
which is formally almost identical to (\ref{eq:subG_X_mean_diff}), with only the subexponential norm replacing the sub-Gaussian norm. However, there is a tacit constraint on (\ref{eq:subE_X_mean_diff_vertex}). Note by definition that $|s|\leq c_\mathrm{E}/\lVert \Delta X\rVert_\mathrm{1E}$, the global infimum in (\ref{eq:thermo_chernoff}) is achieved when $s\lVert \Delta X \rVert_\mathrm{1E}^2/2 = D_\mathrm{KL}(P_0\rVert P_1)/s$, i.e., $s^2=2D_\mathrm{KL}(P_0\rVert P_1)/\lVert \Delta X \rVert_\mathrm{1E}^2$. If this can be satisfied, then by the definition (\ref{eq:subE}) we must have
\begin{eqnarray}\label{eq:subE_valid_for_subG}
D_\mathrm{KL}(P_0\rVert P_1)\leq c_\mathrm{E}^2/2.
\end{eqnarray}
If this is the case, then everything we do to sub-Gaussian variables is the same here. However, this condition is somewhat demanding in general, and if it cannot be satisfied, then the infimum is obtained on the boundary $s = c_\mathrm{E}/\lVert \Delta X\rVert_\mathrm{1E}$, hence
\begin{eqnarray}
|\mathbb{E}_1X - \mathbb{E}_0X| \leq \lVert \Delta X\rVert_\mathrm{1E}\left(\frac{c_\mathrm{E}}{2} + \frac{D_\mathrm{KL}(P_0\rVert P_1)}{c_\mathrm{E}}\right).\notag\\
\end{eqnarray}
One might be tempted to choose a $c_\mathrm{E}$ in our definition (\ref{eq:subE}) so that the condition on $D_\mathrm{KL}(P_0\rVert P_1)$ can be more easily satisfied; however, a bigger $c_\mathrm{E}$ may also result in a bigger $\lVert \Delta X\rVert_\mathrm{E}$, and consequently a less tight bound. Without further information of the dynamics, there seems no reason to expect an optimal choice of $c_\mathrm{E}$.

\subsubsection{Bound for entropy change between two equilibrium states}
Let us also take $X=H_1$. For physical systems, it is also common to see the energy distribution in the form $E^{\alpha-1}e^{-\beta E}$, which is actually the gamma distribution and falls in the class of subexponential distribution. Hence it is also possible that perturbed Hamiltonian $H_1$ is subexponential. Apply the similar analysis as in the sub-Gaussian case, and let $\xi=\text{sgn}(\mathbb{E}_0H_1-\mathbb{E}_1H_1)$, we then obtain the bound on the entropy change that

\begin{eqnarray}
S_1-&&S_0\geq\left(\frac{\xi c_\mathrm{E}}{\lVert\Delta H_1\rVert_{1\mathrm{E}}} - \beta\right)(\mathbb{E}_0H_1-\mathbb{E}_1H_1) - \frac{c_\mathrm{E}^2}{2},
\end{eqnarray}
which is one step ahead of the previous result (\ref{eq:DKL_integral_term}). It is also possible to bound $|S_1-S_0|$, but we omit the result here.

\subsubsection{Subexponential thermodynamic uncertainty
relation}
Next, the thermodynamic uncertainty relation can also be addressed in the subexponential situation. Again, if the condition (\ref{eq:subE_valid_for_subG}) is satisfied, then formally it is straightforward to have for a time-antisymmetric quantity $X$ that
\begin{eqnarray}
2(\mathbb{E}X)^2\leq\lVert \Delta X\rVert_\mathrm{E}^2\Delta S.
\end{eqnarray}
And when the condition is not satisfied,
\begin{eqnarray}
|\mathbb{E}X|\leq\frac{\lVert \Delta X\rVert_\mathrm{E}}{2}\left(\frac{c_\mathrm{E}}{2}+\frac{\Delta S}{c_\mathrm{E}}\right).
\end{eqnarray}

\section{Bounds for Empirical Data}\label{sec:Empirical}

We have so far established theoretical bounds for sub-Gaussian and subexponential variables. However, operationally, there are two issues need to be addressed. First, when analyzing the experimental data, one often uses the sample mean to approximate the expectation. This will inevitably induce some error due to the finite size of data. This issue is particularly relevant when the cost of performing experiments is expensive, and only a limited number of observations can be made. Second, if the sub-Gaussian/subexponential norm and Kullback-Leibler divergence have to be inferred from empirical data (this task is by no means trivial), then it will also bring some uncertainty. There is evidence that thousands of data points are needed to construct an estimate of the Kullback-Leibler divergence between two Gaussian distributions that is close enough to the true value \cite{IEEE:2005}. Hence, the applicability of many theoretical bounds is an important issue that needs further research, especially if these inequalities are expected to hold exactly in a real experimental setting.

In the following, we will first focus on the error related to the finiteness of the sample size, and assume that the norm and divergence or their upper bounds are known. Concentration inequalities that quantify such a kind of error can be established for both sub-Gaussian and subexponential variables. Then we propose a plug-in estimator for inferring the norm from data.

\subsection{Empirical sub-Gaussian bound}\label{subsec:empirical_subG}
Suppose we have $N$ independent, identically distributed data points $X_1,\ldots,X_N$ from experiments. By the law of large numbers, $\hat\mu = \sum_{i=1}^NX_i/N$ converges to $\mathbb{E}X$. However, this process may need a huge number of data especially when the underlying distribution is strongly non-Gaussian. For experimental purposes, here we address the non-asymptotic bound that controls the error incurred by using $\hat\mu$ in our inequalities. That is, if we use $\hat\mu_1$ and $\hat\mu_0$ in (\ref{eq:subG_X_mean_diff}), what will we get? Can we have some sense about the probability $P(|(\hat\mu_1-\hat\mu_0)-(\mathbb{E}_1X-\mathbb{E}_0X)|\geq t)$ for $t\geq 0$? Note $\mathbb{E}(\hat\mu - \mathbb{E}X)=0$, hence $\hat\mu - \mathbb{E}X$ is a centered variable. In fact, $\hat\mu - \mathbb{E}X = \frac{1}{N}\sum_{i=1}^N (X_i-\mathbb{E}X) = \frac{1}{N}\sum_{i=1}^N \Delta X_i$, and
\begin{eqnarray}
\mathbb{E}e^{s(\hat\mu-\mathbb{E}X)} = \mathbb{E}e^{\frac{s}{N}\sum_{i=1}^N\Delta X_i} = \prod_{i=1}^N \mathbb{E}e^{\frac{s}{N}\Delta X_i}.\notag
\end{eqnarray}
Since $\Delta X_i$ is a centered sub-Gaussian variable, then $\mathbb{E}e^{s\Delta X_i/N}\leq e^{s^2\lVert \Delta X \rVert_\mathrm{G}^2/2N^2}$, and we have
\begin{eqnarray}
\mathbb{E}e^{s(\hat\mu-\mathbb{E}X)} \leq e^{s^2\lVert \Delta X \rVert_\mathrm{G}^2/2N},\notag
\end{eqnarray}
hence $\Delta\hat\mu\equiv\hat\mu-\mathbb{E}X$ is a centered sub-Gaussian variable with norm $\lVert \Delta X\rVert_\mathrm{G}/\sqrt{N}$. Also note
\begin{eqnarray}
P(|(\hat\mu_1-\hat\mu_0)-(\mathbb{E}_1X-\mathbb{E}_0X)|\geq t) = P(|\Delta\hat\mu_1 - \Delta\hat\mu_0|\geq t);\notag
\end{eqnarray}
then, for all $s\in\mathbb{R}$,
\begin{eqnarray}
\mathbb{E}e^{s(\Delta\hat\mu_1 - \Delta\hat\mu_0)} &=& \mathbb{E}_1e^{s\Delta\hat\mu_1}\mathbb{E}_0e^{-s\Delta\hat\mu_0}\notag\\
&\leq&e^{s^2\lVert \Delta X\rVert_\mathrm{1G}^2/2N}e^{s^2\lVert \Delta X\rVert_\mathrm{0G}^2/2N}\notag\\
&=&e^{s^2(\lVert \Delta X\rVert_\mathrm{1G}^2+\lVert \Delta X\rVert_\mathrm{0G}^2)/2N},\notag
\end{eqnarray}
hence $\Delta\hat\mu_1 - \Delta\hat\mu_0$ is also sub-Gaussian, with norm $\sqrt{(\lVert \Delta X\rVert_\mathrm{1G}^2+\lVert \Delta X\rVert_\mathrm{0G}^2)/N}$. This leads to the concentration inequality that
\begin{eqnarray}
&&\ P(|(\hat\mu_1-\hat\mu_0)-(\mathbb{E}_1X-\mathbb{E}_0X)|\geq t)\notag\\
=&&\ P(|\Delta\hat\mu_1 - \Delta\hat\mu_0|\geq t)\notag\\
\leq&&\ 2\exp\left(-\frac{Nt^2}{2(\lVert \Delta X\rVert_\mathrm{1G}^2
+\lVert \Delta X\rVert_\mathrm{0G}^2)}\right),\notag
\end{eqnarray}
which is indeed an application of the Hoeffding bound (\ref{eq:Hoeffding}).
Therefore, for all $\delta\in(0,1)$, we have probability at least $1-\delta$ that
\begin{eqnarray}\label{eq:subG_empirical_X_diff}
|\Delta\hat\mu_1 - \Delta\hat\mu_0|\leq \sqrt{\frac{2(\lVert \Delta X\rVert_\mathrm{1G}^2
+\lVert \Delta X\rVert_\mathrm{0G}^2)}{N}\ln\left(\frac{2}{\delta}\right)}.\notag\\
\end{eqnarray}
Put it another way, one has probability at least $1-\delta$ to have a desired error bound $\varepsilon$ on the difference between $\hat\mu_1-\hat\mu_0$ and $\mathbb{E}_1X-\mathbb{E}_0X$, if the number of data points $N$ is on the order of $2(\lVert \Delta X\rVert_\mathrm{1G}^2
+\lVert \Delta X\rVert_\mathrm{0G}^2)\ln(2/\delta)/\varepsilon^2$.

Also, by the triangle inequality and (\ref{eq:subG_X_mean_diff}), we have
\begin{eqnarray}
|\hat\mu_1-\hat\mu_0|
&\leq& |\Delta\hat\mu_1 - \Delta\hat\mu_0|+|\mathbb{E}_1X-\mathbb{E}_0X|\notag\\
&\leq& |\Delta\hat\mu_1 - \Delta\hat\mu_0| + \lVert \Delta X \rVert_\mathrm{1G}\sqrt{2D_\mathrm{KL}(P_0\rVert P_1)}.\notag
\end{eqnarray}
Thus (\ref{eq:subG_empirical_X_diff}) suggests that with probability more than $1-\delta$ we have the experiment-relevant inequality that
\begin{eqnarray}
|\hat\mu_1 - \hat\mu_0|\leq &&\sqrt{\frac{2(\lVert \Delta X\rVert_\mathrm{1G}^2
+\lVert \Delta X\rVert_\mathrm{0G}^2)}{N}\ln\left(\frac{2}{\delta}\right)} \notag\\
&& + \lVert \Delta X \rVert_\mathrm{1G}\sqrt{2D_\mathrm{KL}(P_0\rVert P_1)},
\end{eqnarray}
which sets an upper bound for the absolute difference in sample means at different states. In the case of the sub-Gaussian thermodynamic uncertainty relation (\ref{eq:subG_thermo_inequality}), we have the following bound that with probability more than $1-\delta$
\begin{eqnarray}
|\hat\mu| \leq \lVert \Delta X \rVert_\mathrm{G}\left[ \sqrt{\frac{1}{N}\ln\left(\frac{2}{\delta}\right)} + \sqrt{\frac{1}{2}\Delta S}\right].
\end{eqnarray}

In practice, if there is some way to effectively estimate or upper bound the sub-Gaussian norm and the Kullback-Leibler divergence, our empirical sub-Gaussian result could be applied then.

\subsection{Empirical subexponential bound}
Let us now consider the subexponential case. Following almost exactly the same steps as shown in the sub-Gaussian case, we can establish the concentration bound for such an error. Applying the Bernstein bound (\ref{eq:Bernstein}), we have
\begin{eqnarray}
&&P(|(\hat\mu_1-\hat\mu_0)-(\mathbb{E}_1X-\mathbb{E}_0X)|\geq t)\notag\\
\leq&& 2\exp\left(-\min\left\{
\frac{t^2}{2\lVert\Delta X\rVert_\mathrm{ave}^2},
\frac{t}{\lVert\Delta X\rVert_\mathrm{max}}
\right\}\right),\notag
\end{eqnarray}
where
\begin{eqnarray}
\lVert\Delta X\rVert_\mathrm{ave}^2 &=& \frac{(\lVert \Delta X\rVert_\mathrm{1E}^2
+\lVert \Delta X\rVert_\mathrm{0E}^2)}{N},\ \text{and}\notag\\
\lVert\Delta X\rVert_\mathrm{max} &=& \frac{\max\{\lVert \Delta X\rVert_\mathrm{1E},
\lVert \Delta X\rVert_\mathrm{0E}\}}{2N}.\notag
\end{eqnarray}

Note $\min\{b_1,b_2\}\leq(b_1+b_2)/2$, hence for all $0<\delta<1$ we have the nonasymptotic bound that with probability more than $1-\delta$
\begin{eqnarray}
&&|(\hat\mu_1-\hat\mu_0)-(\mathbb{E}_1X-\mathbb{E}_0X)|\notag\\
\leq&& \lVert\Delta X\rVert_\mathrm{ave}\sqrt{\frac{1}{2}\ln\left(\frac{2}{\delta}\right)}+\lVert\Delta X\rVert_\mathrm{max}\frac{1}{2}\ln\left(\frac{2}{\delta}\right).
\end{eqnarray}
The concentration bound for $|\hat\mu_1 -\hat\mu_0|$ can be similarly obtained as in the sub-Gaussian case. If the subexponential norms can be obtained or nontrivially upper bounded theoretically, then such inequalities can be used to evaluate experimental data.

\subsection{Norm estimation}

In general, we are not aware of any existing work on the inference of the sub-Gaussian/subexponential norm based on data, with a precise uncertainty quantification. Here, based on its definition, we construct a plug-in estimator for the sub-Gaussian norm (the subexponential norm can be treated in a similar way); however, a systematic investigation of its theoretical properties are left for future study. In Appendix \ref{appendix:subGnorm_concentration}, another estimator is constructed based on the concentration inequality for comparison.

Note the definition of sub-Gaussian norm (\ref{eq:subGnorm}), a common practice in statistics is to use the sample mean $\frac{1}{N}\sum_{i=1}^N$ to replace the expectation $\mathbb{E}$. We follow this method to obtain an estimator of $\lVert \Delta X\rVert_\mathrm{G}$, denoted $\hat\sigma_\mathrm{G}$, as
\begin{eqnarray}
\hat\sigma_\mathrm{G} = \inf\left\{ \sigma>0: \frac{1}{N}\sum_{i=1}^Ne^{sx_i}\leq e^{s^2\sigma^2/2}, \forall s \in \mathbb{R} \right\},\notag
\end{eqnarray}
where we have already made $x_i$ a centered variable with respect to the empirical distribution, by subtracting the sample mean $\frac{1}{N}\sum_{i=1}^N X_i$ from the original data $X_i$, hence $\frac{1}{N}\sum_{i=1}^N x_i=0$.
Denote $M(s)=\frac{1}{N}\sum_{i=1}^Ne^{sx_i}$; then $M(s)$ is an asymptotic unbiased estimator of $\mathbb{E}e^{s(X-\mathbb{E}X)}$ for all $s$. It thus reasonable to expect $\hat\sigma_\mathrm{G}$ is an asymptotic unbiased estimator of $\lVert \Delta X\rVert_\mathrm{G}$. By its definition, for each $s$, we have
\begin{eqnarray}
M(s)\leq \exp(\hat\sigma_\mathrm{G}^2s^2/2)\notag,
\end{eqnarray}
which numerically implies
\begin{eqnarray}
\hat\sigma_\mathrm{G}=\min_s \frac{\sqrt{2\ln M(s)}}{|s|}.\notag
\end{eqnarray}
For an estimator for $\lVert\Delta X\rVert_\mathrm{E}$, one can use the same method by noticing the range of $s$ in (\ref{eq:subE}). It is worth stressing that this method provides an point estimation of norm, but the associated confidence interval is not easy to construct at present.

\begin{figure}[tbp]
  \begin{center}
  \epsfig{figure=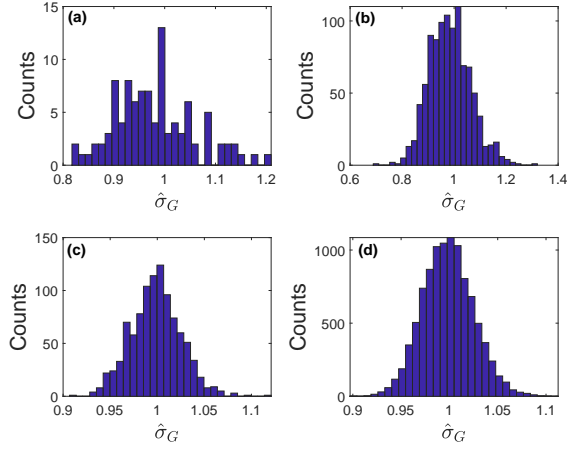,width=\linewidth}\caption{(Color online) Based on the definition, we estimate the sub-Gaussian norm for a simple Gaussian distribution $\mathcal{N}$(0,1), which is known to be $\lVert\Delta X\rVert_\mathrm{G}=1$. We sample $n_d$ data points from the distribution for $n_e$ times and plot the histogram of the estimated sub-Gaussian norm $\hat\sigma_\mathrm{G}$: (a) $n_d=n_e=100$, (b) $n_d=100$, $n_e=1000$, (c) $n_d=n_e=1000$, and (d) $n_d=1000$, $n_e=10000$.} \label{fig:subG_est_def}
  \end{center}
\end{figure}

Let us now consider a simple example. A sample of $n_d$ data points is drawn from the standard Gaussian distribution $\mathcal{N}(0,1)$, and an estimator $\hat\sigma_\mathrm{G}$ is constructed based on it. In this case, we know exactly that $\lVert \Delta X\rVert_\mathrm{G}=1$. We repeat this experiment for $n_e$ times, and we plot the histogram of $\hat\sigma_\mathrm{G}$ in Fig. \ref{fig:subG_est_def}. One can see when $n_d=100$ the relative error can be as high as $20\%$. In many real experiments, the number of points is almost on the same order of magnitude, hence one should be careful when applying $\hat\sigma_\mathrm{G}$ to the corresponding bound (\ref{eq:subG_X_mean_diff}). If a conservative upper bound is crucial, a better idea might be to theoretically bound $\lVert \Delta X\rVert_\mathrm{G}$ like in Sec. \ref{subsec:subG_norm_bound}, otherwise we have to substantially increase the sample size or empirically multiply a safety factor to $\hat\sigma_\mathrm{G}$.

That said, as shown in Fig. \ref{fig:subG_eg}, the bound (\ref{eq:subG_X_mean_diff}) is not necessarily tight, hence an underestimate of $\lVert\Delta X\rVert_\mathrm{G}$ may also work in some cases from a practical perspective. Depending on whether the ``worst case'' or the ``typical case'' matters more in the problem, our confidence could vary in results by using such an estimator $\hat\sigma_\mathrm{G}$ and/or the estimated Kullback-Leibler divergence.

\section{Conclusion and Discussion}\label{sec:Conclusion}

In this work, we have introduced the concepts of sub-Gaussian and subexponential distributions, which seem less known to the statistical physics community. The motivation of our work is that sub-Gaussian and subexponential distributions, as natural generalizations to the Gaussian distribution that facilitates the linear response theory, seem to be particularly relevant to nonlinear response, as hinted by previous experimental and numerical findings. Concrete examples are also provided to show their relevance in statistical physics.

Based on the sub-Gaussian or subexponential norm of a physical quantity, we are able to further develop the theory established by Dechant and Sasa \cite{arXiv:Sasa} for the fluctuation-response relation in general situations out of equilibrium. We refine the bound for the difference in expected values (with respect to two distributions) of an arbitrary variable that falls within the sub-Gaussian or subexponential class. When the distributions considered are about two equilibrium states connected by an external perturbation, we also find a bound that links the entropy difference with the Hamiltonian fluctuation. When the distributions are interpreted as regards to the forward and backward processes, respectively, we obtain a generalized version of the thermodynamic uncertainty relation in each regime. Our results provide universal constraints on the thermodynamical processes without requiring more detailed information of the system in question. But as such information is available, more accurate results with tighter bounds are expected.

Finally, it will be interesting to experimentally test some of our results in real physical systems, which include not only the bounds, but also the plausible transition path from the Gaussian to sub-Gaussian and subexponential distributions as some control parameter is varied. We have also provided non-asymptotic error bounds for the probability that our theoretical bounds hold in an experimental setting. A practical challenge, however, is the estimation of the sub-Gaussian or subexponential norm itself (as well as the Kullback-Leibler divergence). Although plug-in estimators can be constructed, to our knowledge, there is no systematic analysis as to the theoretical properties of these estimators. Alternatively, and maybe practically more preferably, one can use an upper bound of the norm instead, which may be easier to obtain theoretically, as shown in our examples.

\section*{Acknowledgments}
This research was supported in part by the US National Science Foundation under grant HDR:TRIPODS 19-34884. Also, this work cannot be finished without the support from my wife, Dr. Xiaojuan Ma, who sacrifices her own time to take care of the whole family during this COVID-19 pandemic.

\appendix
\section{Properties of sub-Gaussian random variables}\label{appendix:subgaussian}
For a centered sub-Gaussian variable $X$, one can establish the concentration inequality, which is used to bound its tail probability. To this end, note that $P(X\geq t)$ for all $t\geq 0$ and $s>0$ can be bounded as
\begin{eqnarray}
P(X\geq t) = P(e^{sX}\geq e^{st})\leq e^{-st} \mathbb{E}e^{sX},\notag
\end{eqnarray}
where we have used the Markov's inequality in the last step. Then the Chernoff bound states that
\begin{eqnarray}
\ln P(X \geq t) \leq \underset{s>0}{\inf}\{\ln\mathbb{E}e^{sX}-st\},\notag
\end{eqnarray}
combining which with (\ref{eq:subG}) and (\ref{eq:subGnorm}) we obtain that the infimum is achieved at $s=t/\lVert X\rVert_\mathrm{G}^2$, and $P(X\geq t)\leq e^{-t^2/2\lVert X\rVert_\mathrm{G}^2}.$ Similarly, one can bound $P(X\leq -t)$, hence we come to the concentration inequality for the sub-Gaussian variable
\begin{eqnarray}\label{eq:subGCI}
P(|X|\geq t)\leq 2\exp\left({-\frac{t^2}{2\lVert X\rVert_\mathrm{G}^2}}\right),
\end{eqnarray}
which holds for all $t\geq 0$.

The sub-Gaussian property is preserved under addition. Let $X_1,\ldots,X_N$ be independent and centered sub-Gaussian variables with norms $\lVert X_1\rVert_{\mathrm{G}},\ldots,\lVert X_N\rVert_{\mathrm{G}}$, respectively. Let $X=\sum_{i=1}^N a_iX_i$, where $a_i$'s are constants. Then $X$ is also sub-Gaussian, and we can compute its norm by (\ref{eq:subG}) and (\ref{eq:subGnorm}):
\begin{eqnarray}
\mathbb{E}e^{sX} &=& \mathbb{E}e^{s\sum_{i=1}^Na_iX_i} = \prod_{i=1}^N e^{sa_iX_i}\notag\\
&\leq& \prod_{i=1}^N e^{s^2a_i^2\lVert X_i\rVert_{\mathrm{G}}^2/2} = e^{s^2(\sum_{i=1}^Na_i^2\lVert X_i\rVert_{\mathrm{G}}^2)/2}.\notag
\end{eqnarray}
Thus $X$ is sub-Gaussian with $\lVert X\rVert_{\mathrm{G}}^2=\sum_{i=1}^Na_i^2\lVert X_i\rVert_{\mathrm{G}}^2$.
By this result, we have Hoeffding's inequality that
\begin{eqnarray}\label{eq:Hoeffding}
P\left(\left|\frac{1}{N}\sum_{i=1}^N X_i\right| \geq t\right) \leq 2\exp\left(-\frac{N^2t^2}{2\sum_{i=1}^N\lVert X_i\rVert_{\mathrm{G}}^2}\right),\notag\\
\end{eqnarray}
which holds for all $t\geq0$. Furthermore, if these $X_i$'s are identically distributed, then the sample mean $\hat\mu=\sum_{i=1}^NX_i/N$ satisfies $P(|\hat\mu|\geq t)\leq 2e^{-Nt^2/2\lVert X\rVert_\mathrm{G}^2}$.

\section{Properties of subexponential random variables}\label{appendix:subexponential}
One can establish the concentration inequality for subexponential variables in terms of $\lVert X \rVert_\mathrm{E}$. As above, let us start with the Chernoff bound. For $t\geq0$ and $0<s\leq c_\mathrm{E}/\lVert X \rVert_\mathrm{E}$, we have:
\begin{eqnarray}
\ln P(X\geq t)\leq \underset{0<s\leq c_\mathrm{E}/\lVert X \rVert_\mathrm{E}}{\inf}\{\mathbb{E}e^{sX}e^{-st}\}.\notag
\end{eqnarray}
To proceed, let us consider two situations, $t>c_\mathrm{E}\lVert X \rVert_\mathrm{E}$ and $0\leq t\leq c_\mathrm{E}\lVert X \rVert_\mathrm{E}$, since they lead to qualitatively different results.

Noting (\ref{eq:subE}), (\ref{eq:subEnorm}), and the range of $s$, we can see that if $t>c_\mathrm{E}\lVert X \rVert_\mathrm{E}$ then
\begin{eqnarray}
\min_s\exp\left(\frac{1}{2}\lVert X \rVert_\mathrm{E}^2s^2-st\right) = \exp\left(\frac{1}{2}-\frac{c_\mathrm{E}t}{\lVert X \rVert_\mathrm{E}}\right),\notag
\end{eqnarray}
which is achieved at the boundary $s=c_\mathrm{E}/\lVert X \rVert_\mathrm{E}$, and
\begin{eqnarray}
P(X\geq t) &\leq& \exp\left(\frac{1}{2}-\frac{c_\mathrm{E}t}{\lVert X \rVert_\mathrm{E}}\right)\notag\\
&\leq& \exp\left(\frac{\frac{t}{c_\mathrm{E}}-2c_\mathrm{E}t}{2\lVert X \rVert_\mathrm{E}}\right)\notag\\
&=& \exp\left(-\frac{t}{\lVert X \rVert_\mathrm{E}}\right).\notag
\end{eqnarray}
Applying similar arguments to $-X$, then we have for $t>c_\mathrm{E}\lVert X \rVert_\mathrm{E}$ that
\begin{eqnarray}
P(|X|\geq t)\leq 2\exp\left(-\frac{t}{\lVert X \rVert_\mathrm{E}}\right),\notag
\end{eqnarray}
while if $0\leq t\leq c_\mathrm{E}\lVert X \rVert_\mathrm{E}$, then
\begin{eqnarray}
\min_s\exp\left(\frac{1}{2}\lVert X \rVert_\mathrm{E}^2s^2-st\right) = \exp\left(-\frac{t^2}{2\lVert X \rVert_\mathrm{E}^2}\right),\notag
\end{eqnarray}
which is achieved at $s=t/\lVert X\rVert_\mathrm{E}^2$. Hence for $t$ in this range, the situation is the same as in the sub-Gaussian case, and we have $P(|X|\geq t)\leq 2e^{-t^2/2\lVert X \rVert_\mathrm{E}^2}$. Roughly speaking, one can see that for small $t$ deviation from the mean 0, the subexponential variable actually has no difference from a sub-Gaussian variable. The difference manifests itself for large $t$ deviation. Combining the results together, we have the concentration inequality that, for all $t\geq 0$, a centered subexponential variable satisfies
\begin{eqnarray}
P(|X|\geq t) \leq
\left\{
  \begin{array}{ll}
    2\exp\left(-\frac{t^2}{2\lVert X \rVert_\mathrm{E}^2} \right), & \hbox{$0\leq |t|\leq c_\mathrm{E}\lVert X \rVert_\mathrm{E}$;} \\
    2\exp\left(-\frac{t}{\lVert X \rVert_\mathrm{E}} \right), & \hbox{$|t|>c_\mathrm{E}\lVert X \rVert_\mathrm{E}$,}
  \end{array}
\right.\notag
\end{eqnarray}
or in a more compact form,
\begin{eqnarray}
P(|X|\geq t) \leq 2\exp\left(-\min\left\{
\frac{t^2}{2\lVert X \rVert_\mathrm{E}^2},
\frac{t}{\lVert X \rVert_\mathrm{E}}
\right\}\right).\notag
\end{eqnarray}

Slightly different than in the sub-Gaussian case, for $N$ independent centered subexponential variables $X_1,\ldots,X_N$ with norms $\lVert X_1\rVert_{\mathrm{E}},\ldots,\lVert X_N\rVert_{\mathrm{E}}$, there are two relevant norms associated with $X=\sum_{i=1}^Na_iX_i$. As before, for $t\geq 0$ and $0< s\leq c_\mathrm{E}/\max_i\{|a_i|\lVert X_i\rVert_\mathrm{E}\}$, we have
\begin{eqnarray}
P(X\geq t)&\leq& e^{-st}\mathbb{E}e^{\sum_{i=1}^Na_iX_i}
\leq e^{-st}\prod_{i=1}^Ne^{s^2a_i^2\lVert X_i\rVert_\mathrm{E}^2/2}\notag\\
&=&e^{s^2(\sum_{i=1}^Na_i^2\lVert X_i\rVert_\mathrm{E}^2)/2-st}.\notag
\end{eqnarray}
Hence, we can perform an analysis similar to the single variable case. First, if each $X_i$ is in the sub-Gaussian regime for small $t$, then $X$ is sub-Gaussian-like with squared norm $\lVert X \rVert_\mathrm{E\downarrow}^2=\sum_{i=1}^Na_i^2\lVert X_i\rVert_\mathrm{E}^2$. If it is out of this regime, then $X$ is subexponential with squared norm $\lVert X\rVert_{\mathrm{E}\uparrow}^2=N\max_i\{a_i^2\lVert X_i\rVert_\mathrm{E}^2\}$. In the case that $a_i=1/N$, we have the Bernstein's inequality that for all $t\geq 0$:
\begin{eqnarray}\label{eq:Bernstein}
&&P\left(\left|\frac{1}{N}\sum_{i=1}^N X_i\right|\geq t\right)\notag\\
\leq\ && 2\exp\left(-\min\left\{
\frac{N^2t^2}{2\sum_{i=1}^N\lVert X_i \rVert_{\mathrm{E}}^2},
\frac{Nt}{\underset{1\leq i\leq N}{\max}\lVert X_i \rVert_{\mathrm{E}}}
\right\}\right).\notag\\
\end{eqnarray}

\section{Upper bound of the subexponential norm for the centered compression work example}\label{appendix:subE_norm}
The logarithm of the moment generating function of the centered compression work $\Delta W = W-\mathbb{E}W$ is
\begin{eqnarray}
\ln\mathbb{E}e^{s\Delta W} = -\frac{sn}{\beta}-n\ln\left(1-\frac{s}{\beta}\right)\equiv f(s), \notag
\end{eqnarray}
where $s<\beta$ is required. One can easily check $f(0)=0$ and $f'(0)=\mathbb{E}\Delta W=0$. In order to show $\Delta W$ is subexponential, we first calculate $f''(s)$:
\begin{eqnarray}
f''(s) = \frac{n}{(\beta-s)^2}.\notag
\end{eqnarray}
Obviously, $f''(s)>0$, hence $f(s)$ is convex. So if we can upper bound $f''(s)$ by some constant $\sigma^2$ for $|s|\leq c_\mathrm{E}/\sigma$, then it is sufficient to have that, for $0\leq s\leq c_\mathrm{E}/\sigma$,
\begin{eqnarray}
f(s)=\int_0^s\int_0^t f''(u)dudt\leq \int_0^s\int_0^t \sigma^2dudt =\frac{1}{2}\sigma^2s^2.\notag
\end{eqnarray}
Similarly, for $-c_\mathrm{E}/\sigma<s<0$, the same result holds. Then as a result, $\Delta W$ is subexponential. Note $0<f''(s)\leq\sigma^2$ is a sufficient, but not necessary, condition for $f(s)\leq\sigma^2s^2/2$. Hence the subexponential norm is upper bounded by the infimum of such $\sigma$.

Since $f''(s)$ is monotonically increasing in $s$ for $s<\beta$, we thus have
\begin{eqnarray}
\frac{n}{(\beta-s)^2} \leq
\frac{n}{\left(
                \beta - \frac{c_\mathrm{E}}{\sigma}
          \right)^2} \leq \sigma^2,\notag
\end{eqnarray}
which implies
\begin{eqnarray}
\left(\sigma - \frac{c_\mathrm{E}}{\beta}\right)^2 \geq \frac{n}{\beta^2} = \mathrm{var}(W).\notag
\end{eqnarray}
So finally we have
\begin{eqnarray}
\lVert\Delta W\rVert_\mathrm{E} \leq \sigma_\mathrm{E}\equiv\sqrt{\mathrm{var}(W)} + \frac{c_\mathrm{E}}{\beta}.\notag
\end{eqnarray}

Moreover, due to the monotonic property of $f''(s)$, actually for the whole range $-\infty< s\leq c_\mathrm{E}/\sigma_\mathrm{E}$, we have $f(s)\leq \sigma_\mathrm{E}^2s^2/2$. This result is important in analyzing the Jarzynski equality for this model, because
\begin{eqnarray}\label{eq:fss_upper_bound}
&&-\beta(\Delta F -\mathbb{E}W) = \ln\mathbb{E}e^{-\beta(W-\mathbb{E}W)}\notag\\
=&&f(-\beta) = \int_0^{-\beta}\int_0^{t}f''(u)dudt\notag\\
\leq&&\int_0^{-\beta}\int_0^{t}\sigma^2_\mathrm{E}dudt\notag\\
=&&\frac{1}{2}\sigma^2_\mathrm{E}\beta^2.
\end{eqnarray}
Hence we can upper bound $\mathbb{E}W$ as
\begin{eqnarray}
\mathbb{E}W\leq\Delta F + \frac{1}{2}\beta\sigma_\mathrm{E}^2.\notag
\end{eqnarray}

In this specific example, we can actually obtain more by decreasing $\sigma_\mathrm{E}$ to $\sqrt{\mathrm{var}(W)}$ due to the same reasoning since $0\leq f''(s)\leq f''(0) = \mathrm{var}(W)$ for $s\leq 0$. But it is not our focus here, and our aim is to show how to apply the subexponential property, rather than to find a tighter bound of $\mathbb{E}W$.

\section{Derivation of Kullback-Leibler divergence in linear response}\label{appendix:KL_linear_response}
In the linear response regime, $H_1 = H_0 + \varepsilon A$, where $\varepsilon$ is a small parameter. Let us calculate $D_\mathrm{KL}(P_0\rVert P_1)$ first:
\begin{eqnarray}
D_\mathrm{KL}(P_0\rVert P_1) &=& \mathbb{E}_0\ln(P_0/P_1)\notag\\
&=& \mathbb{E}_0[-\beta(H_0-H_1)+\ln(Z_1/Z_0)]\notag\\
&=& \varepsilon\beta\mathbb{E}_0A + \ln(Z_1/Z_0).\notag
\end{eqnarray}
To calculate $\ln(Z_1/Z_0)$, we note
\begin{eqnarray}
Z_1 &=& \int e^{-\beta H_1} = \int e^{-\beta H_0}e^{-\varepsilon\beta A}\notag\\
&\approx& \int e^{-\beta H_0} (1-\varepsilon\beta A + \frac{1}{2}\varepsilon^2\beta^2 A^2),\notag
\end{eqnarray}
hence
\begin{eqnarray}
Z_1/Z_0 &\approx& \int \frac{e^{-\beta H_0}}{Z_0}(1-\varepsilon\beta A + \frac{1}{2}\varepsilon^2\beta^2 A^2)\notag\\
&=& 1 - \varepsilon\beta\mathbb{E}_0A + \frac{1}{2}\varepsilon^2\beta^2 \mathbb{E}_0A^2.\notag
\end{eqnarray}
We thus have
\begin{eqnarray}
\ln(Z_1/Z_0)&\approx&\ln(1 - \varepsilon\beta\mathbb{E}_0A + \frac{1}{2}\varepsilon^2\beta^2 \mathbb{E}_0A^2)\notag\\
&\approx& - \varepsilon\beta\mathbb{E}_0A + \frac{1}{2}\varepsilon^2\beta^2 \mathbb{E}_0A^2 - \frac{1}{2}(\varepsilon\beta\mathbb{E}_0A)^2\notag\\
&=& - \varepsilon\beta\mathbb{E}_0A + \frac{1}{2}\varepsilon^2\beta^2\mathrm{var}_0A.\notag
\end{eqnarray}
Summarizing, we obtain
\begin{eqnarray}
D_\mathrm{KL}(P_0\rVert P_1) \approx \frac{1}{2}\varepsilon^2\beta^2\mathrm{var}_0A.\notag
\end{eqnarray}

Although in general $D_\mathrm{KL}(P_0\rVert P_1)\neq D_\mathrm{KL}(P_1\rVert P_0)$, in this linear region they are the same. To this end, note
\begin{eqnarray}
D_\mathrm{KL}(P_1\rVert P_0)=-\varepsilon\beta\mathbb{E}_1A+\ln(Z_0/Z_1).\notag
\end{eqnarray}
From above, we observe that
\begin{eqnarray}
\ln(Z_0/Z_1) = -\ln(Z_1/Z_0) \approx \varepsilon\beta\mathbb{E}_0A - \frac{1}{2}\varepsilon^2\beta^2\mathrm{var}_0A.\notag
\end{eqnarray}
Also, for an arbitrary quantity $B$, we have
\begin{eqnarray}
\mathbb{E}_1B &=& \int B \frac{e^{-\beta H_1}}{Z_1}\notag\\
&\approx&\int B \frac{e^{-\beta(H_0+\varepsilon A)}}{Z_0(1- \varepsilon\beta\mathbb{E}_0A)}\notag\\
&\approx&(1+\varepsilon\beta\mathbb{E}_0A)\mathbb{E}_0Be^{-\varepsilon\beta A}\notag\\
&\approx& \mathbb{E}_0B-\varepsilon\beta(\mathbb{E}_0AB-\mathbb{E}_0A\mathbb{E}_0B)\notag\\
&=&\mathbb{E}_0B-\varepsilon\beta\mathrm{cov}_0(A,B),\notag
\end{eqnarray}
where $\mathrm{cov}_0(A,B)$ is the correlation between $A$ and $B$ under $P_0$. Letting $A=B$, we have
\begin{eqnarray}
-\varepsilon\beta\mathbb{E}_1A\approx-\varepsilon\beta\mathbb{E}_0A+\varepsilon^2\beta^2\mathrm{var}_0A.\notag
\end{eqnarray}
Combining these results, we find
\begin{eqnarray}
D_\mathrm{KL}(P_1\rVert P_0)\approx \frac{1}{2}\varepsilon^2\beta^2\mathrm{var}_0A,\notag
\end{eqnarray}
and we further conclude that in the linear region Eq. (\ref{eq:DKL_linear_response}) holds.

\section{Upper bound of the sub-Gaussian norm for the Markov chain model}\label{appendix:subGnorm_bound}
By definition, the moment generating function of $r$ is
\begin{eqnarray}
\mathbb{E}e^{sr} = [1-q + qpe^{s/N} + q(1-p)e^{-s/N}]^M,\notag
\end{eqnarray}
from which we know
\begin{eqnarray}
\mathbb{E}r = \left.\frac{d}{ds}\mathbb{E}e^{sr}\right|_{s=0} = \frac{M}{N}q(2p-1).\notag
\end{eqnarray}
Actually, we do not have to calculate the derivative, since on average the number of jumps in the positive direction is $Mqp$, that in the negative direction is $Mq(1-p)$, and each jump changes $1/N$ in the value of $r$, hence $\mathbb{E}r = Mq(2p-1)/N$.

Now we can upper bound the sub-Gaussian norm of $\Delta r = r - \mathbb{E}r$. Notice
\begin{eqnarray}
\ln\mathbb{E}e^{s(r-\mathbb{E}r)}= &&M\ln[1-q + qpe^{s/N} + q(1-p)e^{-s/N}]\notag\\
&&-Mq(2p-1)s/N.\notag
\end{eqnarray}
It is essential to consider the function
\begin{eqnarray}
f(x)=\ln[1-q+qpe^x+q(1-p)e^{-x}] - xq(2p-1),\notag
\end{eqnarray}
where $x=s/N$. We claim that $f(x)$ is convex and $\sigma_\mathrm{G}^2$-smooth, i.e., $f''(x)\leq\sigma_\mathrm{G}^2$. Its global minimum is achieved at $x=0$, where obviously $f(x=0)=0$ . To see these, let us denote
\begin{eqnarray}
c \equiv 1-q,\  a\equiv qpe^x,\ \text{and}\ b\equiv q(1-p)e^{-x}.\notag
\end{eqnarray}
It is easy to see that
\begin{eqnarray}
&& a,b,c\geq 0,\notag\\
&& a'\equiv\frac{da}{dx}=a, \ b'\equiv\frac{db}{dx}=-b, \ \text{and}\ c'\equiv\frac{dc}{dx}=0.\notag
\end{eqnarray}

Let us calculate $f'(x)=df/dx$ first.
\begin{eqnarray}
\frac{df}{dx} = \frac{a-b}{c+a+b} -q(2p-1),\notag
\end{eqnarray}
which we set equal to 0 to get
\begin{eqnarray}
&&a-b - q(2p-1)(c+a+b)=0\notag\\
\Longrightarrow && [1-q(2p-1)]a - [1+q(2p-1)]b - q(2p-1)c = 0.\notag
\end{eqnarray}
Note $1-q(2p-1)\geq 0$ and $1+q(2p-1)\geq 0$ always hold for $0\leq p,q\leq 1$, and 0 is a root of $f'(x)$ since $f'(0)\propto\mathbb{E}\Delta r=0$, hence the above expression can be written as
\begin{eqnarray}
ue^x - ve^{-x} -u+v = 0,\notag
\end{eqnarray}
where $u=[1-q(2p-1)]qp\geq 0$ and $v=[1+q(2p-1)]q(1-p)\geq 0$. Solve the equation to find
\begin{eqnarray}
e^x = -\frac{v}{u}\ \text{or} \ 1,\notag
\end{eqnarray}
but $e^x> 0$, hence $f'(x)$ has one and only one root at $x=0$, which corresponds to the global minimum due to the convexity of $f(x)$ established below.

Next, let us consider $f''(x)=df'(x)/dx$, which turns out to be nonnegative and uniformly upper bounded by some $\sigma_\mathrm{G}^2$:
\begin{eqnarray}
f''(x) &=& \frac{d}{dx}\left[\frac{a-b}{c+a+b}-q(2p-1)\right]\notag\\
&=& \frac{(c+a+b)(a'-b')-(a-b)(c'+a'+b')}{(c+a+b)^2}\notag\\
&=& \frac{c(a+b)+4ab}{(c+a+b)^2}\ \text{[hence $f''(x)\geq 0$]}\notag\\
&=& \frac{c(a+b)}{[c+(a+b)]^2} + \frac{4q^2p(1-p)}{(c+a+b)^2}\ \text{(inserting $a$ and $b$)}\notag\\
&\leq& \frac{1}{4} + \frac{4q^2p(1-p)}{[c+(a+b)]^2}\ \text{[since $c+(a+b)\geq 2\sqrt{c(a+b)}$]}\notag\\
&\leq& \frac{1}{4} + \frac{4q^2p(1-p)}{[2\sqrt{c(a+b)}]^2}=\frac{1}{4} + \frac{4q^2p(1-p)}{4c(a+b)}\notag\\
&\leq& \frac{1}{4} + \frac{4q^2p(1-p)}{4c(2\sqrt{ab})} = \frac{1}{4} + \frac{4q^2p(1-p)}{8cq\sqrt{p(1-p)}}\notag\\
&=& \frac{1}{4} + \frac{1}{2}\frac{q}{1-q}\sqrt{p(1-p)}\ \text{(since $c=1-q$)}\notag\\
&\equiv&\sigma_\mathrm{G}^2.\notag
\end{eqnarray}
Hence $f(x)$ is convex and $\sigma_\mathrm{G}^2$-smooth.

Combining the results that $f(0)=f'(0)=0$, $0\leq f''(x)\leq \sigma_\mathrm{G}^2$, we obtain
\begin{eqnarray}
f(x)\leq \frac{1}{2}\sigma_\mathrm{G}^2 x^2,\notag
\end{eqnarray}
which is equivalent to
\begin{eqnarray}
\ln\mathbb{E}e^{s(r-\mathbb{E}r)} \leq \frac{1}{2}\frac{M}{N^2}\left[\frac{1}{4} + \frac{1}{2}\frac{q}{1-q}\sqrt{p(1-p)}\right]s^2.\notag
\end{eqnarray}
Hence we conclude
\begin{eqnarray}
\lVert\Delta r\rVert_\mathrm{G}^2 \leq \frac{M}{N^2}\left[\frac{1}{4} + \frac{1}{2}\frac{q}{1-q}\sqrt{p(1-p)}\right].\notag
\end{eqnarray}

It is worth stressing that such an upper bound for $\lVert\Delta r\rVert_\mathrm{G}$ is derived for the general model parametrized by probabilities $q$ and $p$. If more detailed information of $p$ and $q$ is known, then we may derive an even more informative bound accordingly.

\begin{figure}[tbp]
  \begin{center}
  \epsfig{figure=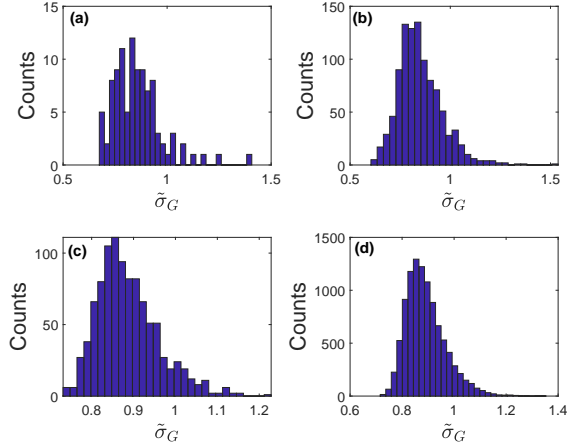,width=\linewidth}\caption{(Color online) Estimation of sub-Gaussian norm based on the concentration inequality. Other settings are the same as in Fig. \ref{fig:subG_est_def}. The true norm equals 1.} \label{fig:subG_est_concentration}
  \end{center}
\end{figure}

Lastly, as a fact check, since $\lVert\Delta r\rVert_\mathrm{G}^2\geq \mathrm{var}(r)$, we must have $\sigma_\mathrm{G}^2 \geq \mathrm{var}(r)=Mq[1-q(2p-1)^2]/N^2$. To this end, we first note that for $0<p,q<1$:
\begin{eqnarray}
&& 1 = 8\times \frac{1}{4} \times\sqrt{\frac{1}{4}}\notag\\
\Longrightarrow\ &&1 \geq 8 q(1-q)\sqrt{p(1-p)}\notag\\
\Longrightarrow\ &&
\frac{1}{2}\frac{q}{1-q}\sqrt{p(1-p)} \geq 4q^2p(1-p).\notag
\end{eqnarray}
On the other hand,
\begin{eqnarray}
&& \frac{1}{4}+4q^2p(1-p) \geq q[1-q(2p-1)^2]\notag\\
\Longleftrightarrow\ && \left(q-\frac{1}{2}\right)^2 \geq 0,\notag
\end{eqnarray}
which is always true. Hence we have confirmed that $\sigma_\mathrm{G}^2\geq\mathrm{var}(r)$.

\section{Estimating sub-Gaussian norm based on the concentration inequality}\label{appendix:subGnorm_concentration}

One might also estimate the sub-Gaussian norm based on the concentration inequality (\ref{eq:subGCI}). However, by doing so, the resulting estimated norm $\hat\sigma_\mathrm{G}$ is biased and typically less than the true value $\lVert \Delta X \rVert_\mathrm{G}$, as is shown in Fig. \ref{fig:subG_est_concentration}, the numerical experimental setting of which is exactly the same as in Fig.\ref{fig:subG_est_def}. Once the data of an experiment are obtained and centered, we first find the data point with maximal absolute value $d$. Then based on the empirical distribution $P_\mathrm{e}$, we calculate $P_\mathrm{e}(|x|\geq t)$ for $t\in(0,d]$. Then we use the following formula to calculate $\tilde\sigma_\mathrm{G}$:
\begin{eqnarray}
\tilde\sigma_\mathrm{G} = \max_t\frac{t}{\sqrt{2\ln\left(\frac{2}{P_\mathrm{e}(|x|\geq t)}\right)}},\notag
\end{eqnarray}
which is essentially the inversion of (\ref{eq:subGCI}). The histogram of $\tilde\sigma_\mathrm{G}$ can then be obtained after many realizations.

Since the sample size is finite and $d$ is finite, but the true Gaussian distribution is unbounded, hence one expects that typically $P_\mathrm{e}(|x|\geq t)\leq P(|\Delta X|\geq t)$, and it is not surprising that oftentimes $\tilde\sigma_\mathrm{G}$ underestimates $\lVert \Delta X \rVert_\mathrm{G}$.


\begin{references}

\bibitem{Book:Landau}
L. D. Landau, and E. M. Lifshitz, \emph{Statistical Physics, Part I}, 3rd ed. (Pergamon Press, New York, 1980).

\bibitem{Book:Kubo}
R. Kubo, M. Toda, and N. Hashitsume, \emph{Statistical Physics II: Nonequilibrium Statistical Mechanics}, 2nd ed. (Springer, New York, 1998).

\bibitem{PhysRep:2008}
U. M. B. Marconi, A. Puglisi, L. Rondoni, and A. Vulpiani, Fluctuation-dissipation: Response theory in statistical physics, Phys. Rep. {\bf 461}, 111 (2008).

\bibitem{PRL:Jarzynski}
C. Jarzynski, Nonequilibrium Equality for Free Energy Differences, Phys. Rev. Lett. {\bf 78}, 2690 (1997).

\bibitem{RPP:Seifert}
U. Seifert, Stochastic thermodynamics, fluctuation theorems and molecular machines, Rep. Prog. Phys. {\bf 75}, 126001 (2012).

\bibitem{Science:2002}
J. Liphardt, S. Dumont, S. B. Smith, I. Tinoco Jr., and C. Bustamante, Equilibrium Information from Nonequilibrium Measurements in an Experimental Test of Jarzynski Equality, Science {\bf 296}, 1832 (2002).

\bibitem{Nature:2005}
D. Collin, F. Ritort, C. Jarzynski, S. B. Smith, I. Tinoco Jr., and C. Bustamante, Verification of the Crooks fluctuation theorem and recovery of RNA folding free energies, Nature (London) {\bf 437}, 231 (2005).

\bibitem{PRL:VdB2006}
B. Cleuren, C. Van den Broeck, and R. Kawai, Fluctuation and Dissipation of Work in a Joule Experiment, Phys. Rev. Lett. {\bf 96}, 050601 (2006).

\bibitem{PRL:Seifert2006}
V. Blickle, T. Speck, L. Helden, U. Seifert, and C. Bechinger, Thermodynamics of a Colloidal Particle in a Time-Dependent Nonharmonic Potential, Phys. Rev. Lett. {\bf 96}, 070603 (2006).

\bibitem{PRL:2009}
M. Bonaldi \emph{et al.}, Nonequilibrium Steady-State Fluctuations in Actively Cooled Resonators, Phys. Rev. Lett. {\bf 103}, 010601 (2009).

\bibitem{PRE:2009}
A. Engel, Asymptotics of work distributions in nonequilibrium systems, Phys. Rev. E {\bf 80}, 021120 (2009).

\bibitem{arXiv:H2020}
P. Chvosta, D. Lips, V. Holubec, A. Ryabov, and P. Maass, Statistics of work performed by optical tweezers with general time-variation of their stiffness, J. Phys. A: Math. Theor. {\bf 53}, 275001 (2020).

\bibitem{PRL:2019}
H. J. D. Miller, M. Scandi, J. Anders, and M. Perarnau-Llobet, Work Fluctuations in Slow Processes: Quantum Signatures and Optimal Control, Phys. Rev. Lett. {\bf 123}, 230603 (2019).

\bibitem{arXiv:Q2020}
Z. Fei and H. T. Quan, Nonequilibrium Green's Function's Approach to the Calculation of Work Statistics, Phys. Rev. Lett. {\bf 124}, 240603 (2020).

\bibitem{Book:Martin}
M. J. Wainwright, \emph{High-Dimensional Statistics: A Non-Asymptotic Viewpoint} (Cambridge University Press, New York, 2019).

\bibitem{Book:Roman}
R. Vershynin, \emph{High-Dimensional Probability: An Introduction with Applications in Data Science} (Cambridge University Press, New York, 2018).

\bibitem{note}
$P(X\leq -x)=\int_{-\infty}^{-x} P(u)du
= \int_{x}^{\infty} P(u)e^{-u}du \leq e^{-x}\int_{-\infty}^{\infty}P(u)du
= e^{-x}$.

\bibitem{PRE:CJ}
G. E. Crooks and C. Jarzynski, Work distribution for the adiabatic compression of a dilute and interacting classical gas, Phys. Rev. E {\bf 75}, 021116 (2007).

\bibitem{note:f}
It is trivial that $f(0)=f'(0)=0$. The fact that $f''(s)>0$ can be seen by noticing $f''(s)= \mathbb{E}\left[(W-\mathbb{E}W)^2\frac{e^{s(W-\mathbb{E}W)}}{\mathbb{E}e^{s(W-\mathbb{E}W)}}\right] - \left\{\mathbb{E}\left[(W-\mathbb{E}W)\frac{e^{s(W-\mathbb{E}W)}}{\mathbb{E}e^{s(W-\mathbb{E}W)}}\right]\right\}^2$, which is equivalent to the variance of $W$ as if $W\sim P'$, where $P'$ is not the real distribution $P$ of $W$, but it is related to $P$ with the Radon-Nikodym derivative being $dP'/dP=\frac{e^{s(W-\mathbb{E}W)}}{\mathbb{E}e^{s(W-\mathbb{E}W)}}$. Hence $f''(s)>0$ if $W$ is not a constant.

\bibitem{arXiv:Sasa}
A. Dechant and S.-i. Sasa, Fluctuation-response inequality out of equilibrium, Proc. Natl. Acad. Sci. U.S.A. {\bf 117}, 6430 (2020).

\bibitem{PRL:TUR01}
A. C. Barato and U. Seifert, Thermodynamic Uncertainty Relation for Biomolecular Processes, Phys. Rev. Lett. {\bf 114}, 158101 (2015).

\bibitem{PRL:TUR02}
T. R. Gingrich, J. M. Horowitz, N. Perunov, and J. England, Dissipation Bounds All Steady-State Current Fluctuations, Phys. Rev. Lett. {\bf 116}, 120601 (2016).

\bibitem{note:subGnorm}
One can prove that $\lVert X\rVert_\mathrm{G}$ is indeed a norm on a common probability space of centered sub-Gaussian random variables. It is easy to see (i) $\lVert X\rVert_\mathrm{G}=0\Rightarrow X=0$ almost surely, since $0\leq \mathbb{E}X^2=\mathrm{var}(X)\leq\lVert X\rVert_\mathrm{G}^2=0$, (ii) $\lVert aX\rVert_\mathrm{G}=|a|\lVert X\rVert_\mathrm{G}$ for $a\in\mathbb{R}$ by definition, and finally (iii) $\lVert X+Y\rVert_\mathrm{G}\leq \lVert X\rVert_\mathrm{G} + \lVert Y\rVert_\mathrm{G}$ by invoking the H\"{o}lder inequality that $\mathbb{E}e^{s(X+Y)}\leq (\mathbb{E}e^{psX})^{1/p}(\mathbb{E}e^{qsY})^{1/q}$ with $p,q\geq 1$ and $p^{-1}+q^{-1}=1$. Pick $p=1+\lVert Y\rVert_\mathrm{G}/\lVert X\rVert_\mathrm{G}$ to obtain $\mathbb{E}e^{s(X+Y)}\leq e^{(\lVert X\rVert_\mathrm{G}+\lVert Y\rVert_\mathrm{G})^2s^2/2}$ for all $s\in\mathbb{R}$, completing the proof.

\bibitem{note:subEnorm}
One can similarly prove $\lVert X\rVert_\mathrm{E}$ is a norm on a common probability space of centered subexponential random variables as in the sub-Gaussian case \cite{note:subGnorm}. The first two points are easy to see. For point (iii), note that in order to apply the H\"{o}lder inequality, we require $\mathbb{E}e^{s(X+Y)}\leq e^{(\lVert X\rVert_\mathrm{E}+\lVert Y\rVert_\mathrm{E})^2s^2/2}$ to be valid for $|s|\leq c_\mathrm{E}/(\lVert X\rVert_\mathrm{E}+\lVert Y\rVert_\mathrm{E})$, which is satisfied since $c_\mathrm{E}/(\lVert X\rVert_\mathrm{E}+\lVert Y\rVert_\mathrm{E}) \leq \min\{c_\mathrm{E}/\lVert X\rVert_\mathrm{E}, c_\mathrm{E}/\lVert Y\rVert_\mathrm{E}\}$. Hence we have $\lVert X+Y\rVert_\mathrm{E}\leq \lVert X\rVert_\mathrm{E} + \lVert Y\rVert_\mathrm{E}$.

\bibitem{Book:Villani}
C. Villani, \emph{Optimal Transport: Old and New} (Springer, Berlin, 2009).

\bibitem{note:W1}
$W_1$ is given by $W_1(P_0,P_1)=\sup_{f\in\mathrm{Lip}_1}\mathbb{E}_0f-\mathbb{E}_1f$, and $f\in\mathrm{Lip}_1$ means $f$ is 1-Lipschtz: $|f(\omega)-f(\omega')|\leq \lVert\omega-\omega'\rVert$ with $\lVert\cdot\rVert$ being some proper norm on the space $\Omega$. $W_1$ defines a metric on the probability space $\mathcal{P}$, which is nonnegative, equal to 0 only when $P_0=P_1$, symmetric in $P_0$ and $P_1$, and satisfies the triangle inequality.

\bibitem{arXiv:Dechant}
A. Dechant and Y. Sakurai, Thermodynamic interpretation of Wasserstein distance, arXiv:1912.08405.

\bibitem{arXiv:TVV}
T. Van Vu and Y. Hasegawa, Geometrical Bounds of the Irreversibility in Markovian Systems, arXiv:2005.02871.

\bibitem{PRL:Seifert2005}
U. Seifert, Entropy Production along a Stochastic Trajectory and an Integral Fluctuation Theorem, Phys. Rev. Lett. {\bf 95}, 040602 (2005).

\bibitem{PRL:VdB2007}
R. Kawai, J. M. R. Parrondo, and C. Van den Broeck, Dissipation: The Phase-Space Perspective, Phys. Rev. Lett. {\bf 98}, 080602 (2007).

\bibitem{PRL:TURFT}
Y. Hasegawa and T. Van Vu, Fluctuation Theorem Uncertainty Relation, Phys. Rev. Lett. {\bf 123}, 110602 (2019).

\bibitem{IEEE:2005}
Q. Wang, S. R. Kulkarni, and S. Verd\'{u}, Divergence Estimation of Continuous Distributions Based on Data-Dependent Partitions, IEEE Trans. Inf. Theory {\bf 51}, 3064 (2005).

\end{references}
\end{document}